\documentclass[usenatbib,letterpaper]{mn2e}
\usepackage[totalwidth=480pt,totalheight=680pt,twoside=false,voffset=.5cm]{geometry}

\usepackage{graphics,graphicx,epsfig,ulem,multirow,appendix,lscape,amsmath,amssymb,color}
\DeclareMathOperator{\sinc}{sinc}

\begin{document}

\title[Effect of an unseen mass on derived companion orbits]
  {Imaged sub-stellar companions: not as eccentric as they appear? The effect of an unseen inner mass on derived orbits}
\author[T. D. Pearce, M. C. Wyatt \& G. M. Kennedy]
  {Tim D. Pearce\thanks{tdpearce@ast.cam.ac.uk},
	Mark C. Wyatt and Grant M. Kennedy\\
  Institute of Astronomy, University of Cambridge, Madingley Road, Cambridge, CB3 0HA, UK}
\date{Released 2002 Xxxxx XX}

\pagerange{\pageref{firstpage}--\pageref{lastpage}} \pubyear{2002}

\def\LaTeX{L\kern-.36em\raise.3ex\hbox{a}\kern-.15em
    T\kern-.1667em\lower.7ex\hbox{E}\kern-.125emX}

\newtheorem{theorem}{Theorem}[section]

\label{firstpage}

\maketitle


\begin{abstract}              
\noindent Increasing numbers of sub-stellar companions are now being discovered via direct imaging. Orbital elements for some of these objects have been derived using star--companion astrometry, and several of these appear to have eccentricities significantly greater than zero. We show that stellar motion caused by an undetected inner body may result in the companion elements derived in such a way being incorrect, which could lead to an overestimation of the eccentricity. The magnitude of this effect is quantified in several regimes and we derive the maximum eccentricity error a third body could introduce in a general form, which may be easily applied to any imaged system. Criteria for identifying systems potentially susceptible to this scenario are presented, and we find that around half of the planets/companion brown dwarfs currently imaged could be liable to these errors when their orbital elements are derived. In particular, this effect could be relevant for systems within 100 pc with companions at \textgreater 50 AU, if they also harbour an unseen $\sim 10$ Jupiter mass object at \textgreater 10 AU. We use the Fomalhaut system as an example and show that a 10\% error could be induced on the planet's eccentricity by an observationally allowed inner mass, which is similar in size to the current error from astrometry.

\end{abstract}



\section{Introduction}
\label{section: Introduction}

\noindent The past two decades have witnessed the birth of direct imaging as a technique to detect sub-stellar companions, with the first discovery of an orbiting brown dwarf \citep{Nakajima95} and later a giant planet \citep{Chauvin04} via this method. Many more potential companions have since been imaged around other stars \citep{exoplanet.eu}, with the method favouring large objects at wide separations from their hosts. In addition the detection of orbital motion between imaging epochs has allowed constraints to be placed on some companion orbits (e.g. \citealt{Soummer11, Chauvin12}), with several appearing to have eccentricities significantly greater than zero (e.g. \citealt{Neuhauser10, Kalas13}).

Planet formation models generally favour the production of low eccentricity companions, as any eccentricity excitations are quickly damped by the gas disk early on \citep{Lissauer93}. Gravitational instability is also thought to initially form protoplanets on low eccentricity orbits \citep{Boss11}. Therefore the existence of eccentric companions imply some further process occurs beyond formation, which could be planet-planet scattering \citep{Gladman93, Marzari02}, 3+ body effects such as secular perturbations \citep{Lee03}, stellar flybys \citep{Malmberg11} or even planet mergers \citep{Lin97} to name a few mechanisms. An accurate measure of eccentricity is very important for a dynamical understanding of these systems, and an overestimation of this quantity could result in an incorrect understanding of system evolution. A potential source of systematic overestimation of eccentricity in imaged systems is the subject of this paper.

Orbital elements of extrasolar companions detected via any method are generally derived in an astrocentric frame (relative to the star) assuming no other bodies in the system. However if an undetected third mass were also present then it would induce a stellar motion about the system barycentre, which could lead to the companion elements derived in this way being incorrect. This  effect has already been examined for radial velocity (RV) detections; \cite{Rodigas09} showed that 10-20\% of RV companions with eccentricities of 0.1-0.4 could actually be on circular orbits with an error introduced by an undetected outer companion, and \cite{Wittenmyer13} identified several moderately eccentric single planet systems that could be better fitted by two low eccentricity planets. However a similar effect has not been considered for wide separation companions detected by imaging, where an additional mass could lie interior to this object and perturb the stellar motion.

Indeed, the existence of an unseen massive object interior to an imaged companion is often suggested if the latter is on a large eccentric orbit, as the inner mass may be required to scatter the observed object out to such a wide separation \citep{Kalas13}. In addition, long-term RV trends (e.g. \citealt{Segransan11}) and micro-lensed planets \citep{Gaudi12} in some systems suggest that companions at $\sim 10$ AU may be common, and these objects could have significant masses yet still remain unseen due to limitations in detection methods. High contrast imaging at these separations is difficult and whilst RV surveys have excelled in locating short period companions, detectable planets in Jovian type orbits remain elusive. Furthermore the precision of RV measurements is significantly reduced when applied to young stars due to stellar activity, yet it is in these systems that outer companions are easiest to detect with imaging. This is highlighted by the case of $\beta$ Pictoris, which shows that massive objects ($\sim 8 {\rm M_J}$, where ${\rm M_J}$ is the mass of Jupiter) may exist around such stars yet evade RV detection \citep{Lagrange09,Lagrange12}.

If an unseen massive object existed in a system with a wide separation imaged companion, then this companion could in fact orbit the star--inner object barycentre. The motion of the star about this barycentre would then cause the astrocentric elements of the imaged companion to vary with a period similar to that of the inner object (e.g. \citealt{Morbidelli02}), and hence its observationally derived orbital elements would be incorrect. In this work we examine the effect of such a scenario on the derived eccentricity of the outermost object, which could be overestimated if an inner companion were present. 

The layout of this paper is as follows. Sections \ref{sec: circular orbit} and \ref{sec: eccentric orbit} describe the theory work. In Section \ref{sec: circular orbit} we consider the case where the observed companion is on a circular barycentric orbit, in order to find the minimum mass of an unseen inner object required to make the outer body appear eccentric. We then generalise this to an eccentric outer companion in Section \ref{sec: eccentric orbit} to find the maximum error in eccentricity which could be induced by an inner object. We suggest criteria to identify systems potentially susceptible to these scenarios in Section \ref{sec: applicability}, and Section \ref{sec: how to use} provides a step by step method which may be used to evaluate the magnitude of this effect for a given system. We apply this method to Fomalhaut and several other systems as examples. We remark on the detectability of an inner mass in Section \ref{sec: detectability}, and discussion and conclusions are given in Sections \ref{sec: discussion} and \ref{sec: conclusions}.


\section{Outer object on circular orbit}
\label{sec: circular orbit} 
\subsection{Negligible time between observations}
\label{sec: case 1} 

\noindent Firstly we investigate how an object on a circular barycentric orbit may be given an apparent astrocentric eccentricity by an unseen inner companion. We assume that the outer object is small compared to the star, and is sufficiently distant that it undergoes two-body motion about the star-inner mass barycentre. We also make the initial assumption that the astrocentric position and velocity of the observed companion are both known at a single epoch (i.e. the time between observations required to derive the velocity is negligible compared to the inner object period), which will later be relaxed. Finally the three body system is assumed to be coplanar, though we will later show that any mutual inclination reduces the effect of an inner mass on the astrocentric elements of the outer body.

\begin{figure}
  \centering
      \includegraphics[width=8cm]{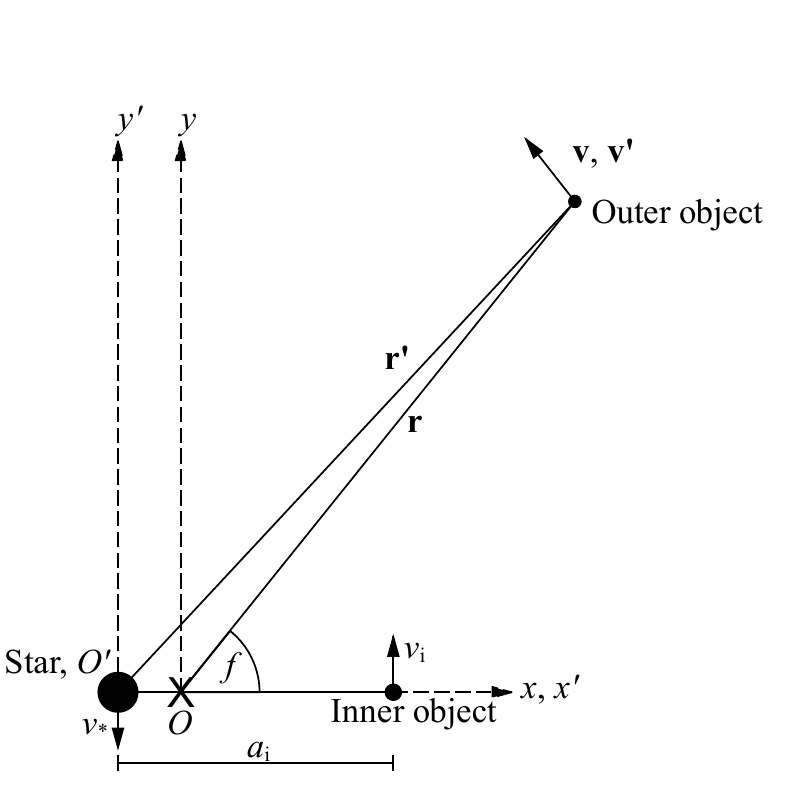}
  \caption{Set-up for Section \ref{sec: case 1}. The star and inner object form a
circular binary of separation $a_{\rm i}$ orbiting their barycentre $O$. The
barycentric frame is centred on $O$, with the inner object on the positive $x$
axis and the star moving in the negative $y$ direction. The outer object is on a
circular orbit about $O$ in the this frame, with a true anomaly $f$ defined
from the $x$ axis. The astrocentric (primed) frame is centred on the star
($O'$) with the $x'$ and $y'$ axes parallel to $x$ and $y$ respectively.}
  \label{fig: case1 setup}
\end{figure}

The set-up of this problem is shown on Figure \ref{fig: case1 setup}. The star and inner object form a circular binary, and the observed companion orbits the binary barycentre. At the moment of observation the barycentric coordinate system is defined to be aligned with the binary separation vector, and the observed object has a true anomaly $f$ in this frame. We also define an astrocentric coordinate system centred on the star, with the axes parallel to those in the barycentric frame. The astrocentric position of the observed companion $\mathbf{r'}$ is therefore given by $\mathbf{r} - \mathbf{r_*}$, its barycentric position minus that of the star, and its astrocentric velocity $\mathbf{v'}$ is given by a similar expression. We will use primes to denote astrocentric parameters for the remainder of the paper, and the subscript i will be used to identify parameters associated with the inner object.

We can show that $\mathbf{r'}$ is given by

\begin{equation}
 \mathbf{r'} = a \left( \begin{array}{c}
      \cos f \\
      \sin f   \end{array} \right) + \mu a_{\rm i} \left(
\begin{array}{c}
      1 \\
      0   \end{array} \right),
 \label{eq: r'}
\end{equation}

\noindent where $a$ denotes the barycentric outer object semi-major axis, $a_{\rm i}$ is the binary separation, and $\mu \equiv m_{\rm i}/(m_* + m_{\rm i})$ where $m_{\rm i}$ and $m_{*}$ are the masses of the inner object and star respectively. Additionally the velocity is

\begin{equation}
 \mathbf{v'} = \sqrt{\frac{GM}{a}} \left( \begin{array}{c}
      - \sin f \\
      \cos f   \end{array} \right) + \mu \sqrt{\frac{GM}{a_{\rm i}}} \left( \begin{array}{c}
      0 \\
      1   \end{array} \right)
\label{eq: v'}
\end{equation}

\noindent where $M \equiv m_* + m_{\rm i}$. To simplify the following we introduce the parameter $\alpha \equiv a_{\rm i} / a$ that, along with $\mu$, contains all the information required to calculate the astrocentric coordinates. The fractional difference between the astrocentric and barycentric radii, $ \delta r / r \equiv (r' - r) / r$, therefore has a maximum absolute value of $\mu \alpha$. Similarly, $\delta v / v = \mu / \sqrt{\alpha}$ at its maximum value. As we only consider unseen companions interior to the observed object, $\alpha < 1$ so the difference between the astrocentric and barycentric radii of the outer object is small whilst the velocity difference may be large. For example, a $0.01 m_*$ object orbiting at $\alpha = 0.1$ would give the observed companion maximum $\delta v / v$ and $\delta r / r$ values of 0.03 and 0.001 respectively.

\begin{figure}
  \centering
      \includegraphics[width=8cm]{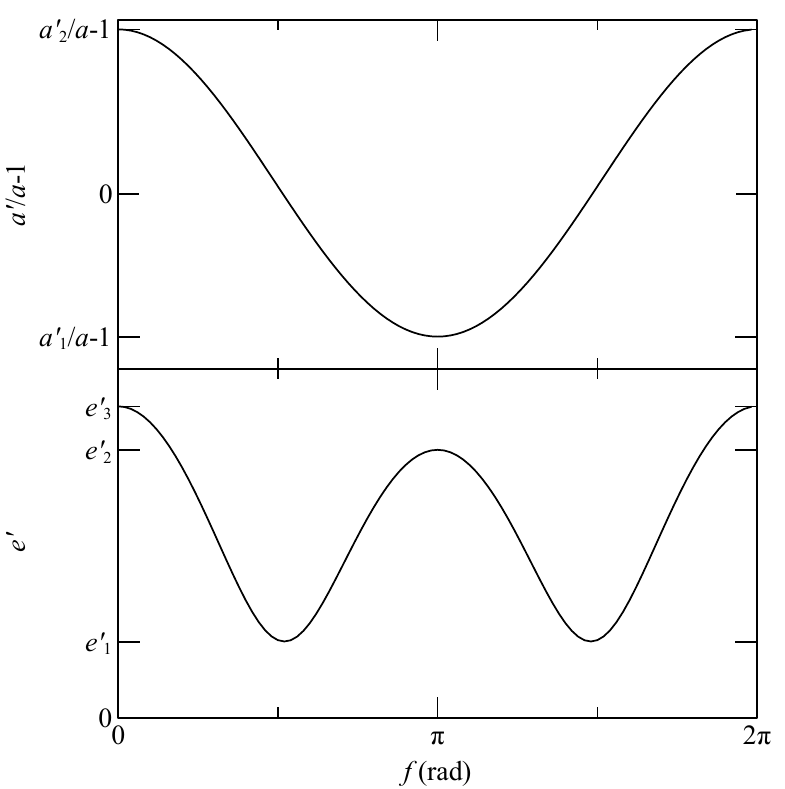}
  \caption{Astrocentric semi-major axis and eccentricity of an outer object on a circular barycentric orbit, in the case where the time between observations is small. The turning points of both elements are denoted by subscripts, and may be well approximated by substituting $f = 0$, $\pi/2$ and $\pi$ into Equations \ref{eq: a' 1st order} and \ref{eq: e' 1st order}. Note that the astrocentric eccentricity is never zero, i.e. $e'_1 > 0$. The plots depend only on $\mu$ and $\alpha$, and are qualitatively the same for all parameters.}
  \label{fig: osculating elements}
\end{figure}

We convert the astrocentric Cartesian coordinates $\mathbf{r'}$ and $\mathbf{v'}$ into Keplerian orbital elements, and the resulting semi-major axis and eccentricity are shown as functions of true anomaly $f$ on Figure \ref{fig: osculating elements}. The plotted functions are quite cumbersome, but to give the reader a feel for their behaviour we simplify them to the following first order approximations. When $\alpha$ is small, $\delta v / v \gg \delta r / r$ and the behaviour of the osculating elements are completely dominated by the velocity shift. In this case the semi-major axis and eccentricity reduce (to first order in $\mu$) to

\begin{equation}
 \frac{a'-a}{a} \approx \mu \left[1 + \frac{2}{\sqrt{\alpha}} \cos f \right]
\label{eq: a' 1st order}
\end{equation}

\noindent and

\begin{equation}
 e' \approx \mu \left[1 + \frac{4}{\sqrt{\alpha}} \cos f + \frac{1}{\alpha} \left(1 + 3 \cos^2 f \right)\right]^\frac{1}{2},
\label{eq: e' 1st order}
\end{equation}

\noindent which are roughly of order $\mu / \sqrt{\alpha}$, the same as the velocity shift. These equations provide a very good fit to the full functions, and hence the turning points on Figure \ref{fig: osculating elements} may be estimated by substituting $f = 0$, $\pi/2$ and $\pi$ into the above. For example if a companion on a 50 AU circular orbit were observed about a solar type star, an undetected $10 {\rm M_J}$ object at 1 AU would cause the observed companion's apparent semi-major axis to vary between $43.5-57.5$ AU and its eccentricity to oscillate between 0.07 and 0.15 with a sub maximum at 0.13.

Note that as $\alpha$ approaches unity, additional $\alpha$ terms caused by the radial shift $\delta r / r$ are no longer negligible in comparison to $1/\sqrt{\alpha}$, so Equations \ref{eq: a' 1st order} and \ref{eq: e' 1st order} no longer hold. Regardless as we assume the outer object undergoes two body motion about the barycentre, the model is invalid in this regime due to three body interactions. However we do not consider such a scenario due to the nature of the problem; as $\alpha \rightarrow 1$ the mass of this object would have to be large to have any effect and should therefore be detectable. We do not consider $\alpha > 1$ for the same reason, and additionally the detected companion would be unlikely to orbit the barycentre in this case.

\begin{figure}
  \centering
      \includegraphics[width=8cm]{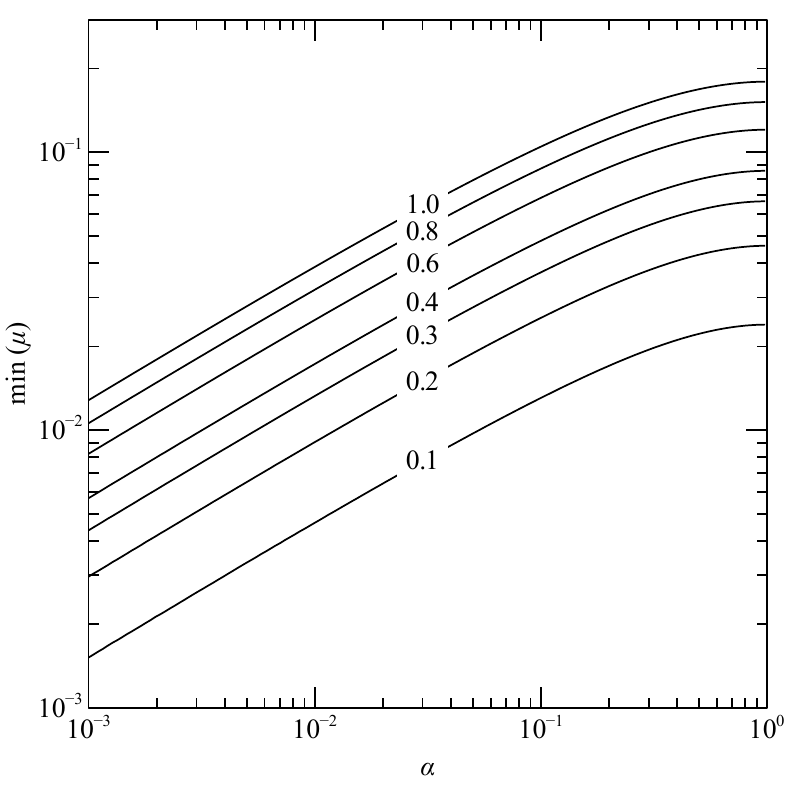}
  \caption{Minimum $\mu$ required to give an outer object on a circular orbit an apparent eccentricity $e'$, if the time between observations is much smaller than the inner object period (see Section \ref{subsec: case 2} if this is not the case). Each line shows a different $e'$. Note the change in behaviour as $\alpha$ approaches unity as described in the text. This plot is independent of $m_*$ and $a$.}
  \label{fig: case1 min mu}
\end{figure}

The maximum values of $\delta r$ and $\delta v$ occur when $f = 0$, i.e. all bodies are aligned, with the star farthest from the outer object. Here the stellar motion opposes the motion of the observed companion, and therefore $\delta v$ is maximised. We can differentiate the full equations for $a'$ and $e'$ and show that these elements are also maximum here. Therefore by substituting $f = 0$ into the full equation for $e'$, we find an upper bound on $e'$ for each combination of $\mu$ and $\alpha$. This equation may be rearranged to find the minimum value of $\mu$ (as a function of $\alpha$) required to give the outer object an apparent astrocentric eccentricity $e'$. The resulting expression contains terms up to high orders in $\mu$, however it may be approximated to better than $5\%$ accuracy by discarding terms greater than second order and multiplying by an empirical factor $F(e')$ to account for higher order terms. This yields the equation

\begin{equation}
\mu \gtrsim F(e') e'\left[1 + 4\left(\frac{1}{\alpha} + \frac{1}{\sqrt{\alpha}} + \sqrt{\alpha} + \alpha^2 \right) + 2\alpha \right]^{-\frac{1}{2}},
\label{eq: case1 min mu}
\end{equation}

\noindent where $F(e') \equiv (1+0.3 e')^{-1}$ is the empirically determined factor. Without this factor the above formula overestimates the minimum value of $\mu$ by $\sim 25\%$ for high values of $e'$. The minimum value of $\mu$ is therefore only dependent on $\alpha$ and the observed astrocentric eccentricity, so is applicable to all systems. Figure \ref{fig: case1 min mu} shows this minimum mass as a function of $\alpha$; the contours were calculated using the full formalism rather than Equation \ref{eq: case1 min mu}, but are well approximated by the latter. It is clear that the inner mass required to give the outer object a significant apparent eccentricity is generally large, typically in the giant planet to brown/red dwarf regime for a solar type star. 

As $\delta r / r$ is small the approximation $a \approx r'$ is generally very good, so this may be used to derive $a_{\rm i}$ from $\alpha$. Also as the minimum value of $e'$ is non-zero, we could progress in the same way as above to derive an upper limit on inner object mass and thus bound $\mu$ in $\alpha$ space. However this limit is not provided as the value is high (such an object would be identifiable using other methods, such as spectroscopy or imaging), so a better upper bound will be given by observational limits.

We have assumed that the three body system is coplanar to derive the above bounds. If this condition is relaxed, we find that any mutual inclination reduces the difference between the two sets of outer object elements. This is to be expected; as noted above, the effect is maximised when the velocity shift $\delta v$ is greatest, i.e. when the direction of the stellar motion opposes that of the outer body. Mutual inclination reduces the stellar velocity component in the outer companion's orbital plane, and hence lowers the velocity shift and thus its effect on the latter's elements. Therefore the value of $\mu$ derived using Equation \ref{eq: case1 min mu} will always be the minimum even if mutually inclined orbits are considered.


\subsection{Non-negligible time between observations}
\label{subsec: case 2}

\begin{figure}
\noindent
\includegraphics[width=8cm]{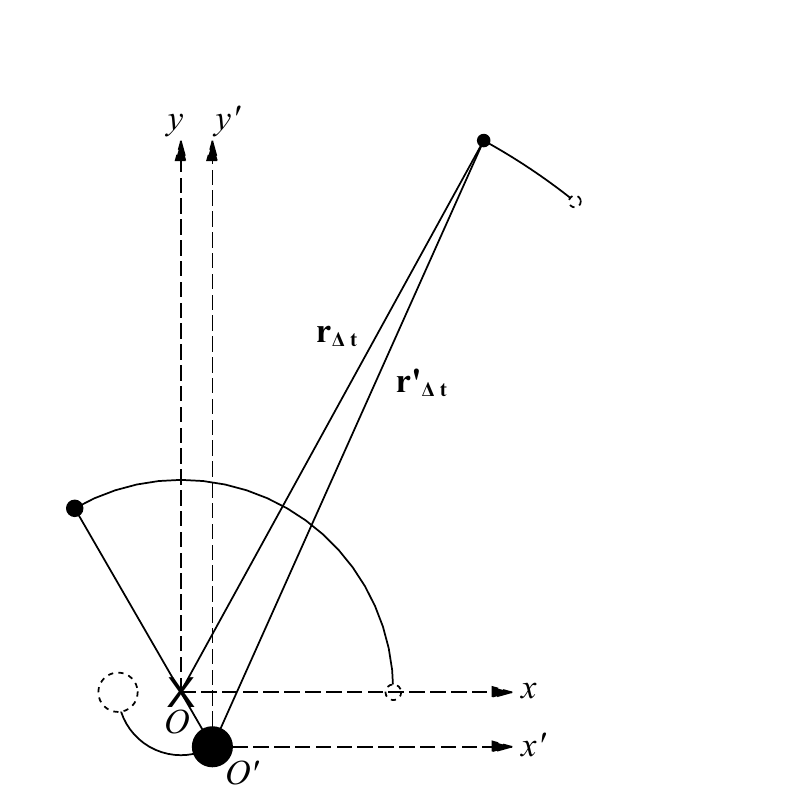}
\caption{Triple system at a time $\Delta t$ after the first observation. Note that at both epochs the barycentric coordinate system is defined with respect to the inner binary position at $t = 0$.}
\label{fig: case2 setup} 
\end{figure}

Figure \ref{fig: case1 min mu} suggests that the minimum $\mu$ required to give a circular companion an apparent eccentricity may always be reduced by placing the inner object ever closer to the star. Unfortunately there is a problem encountered in this regime, as the above assumes that the astrocentric coordinates of the outer object are known instantaneously. In reality the velocity is derived by taking (at least) two images at two different epochs, and between these epochs the inner binary has also progressed about its orbit, as shown on Figure \ref{fig: case2 setup}. The effect of this motion will be significant if the time between observations is of the order of the inner binary period $T_{\rm i}$ or greater, so is most important for inner objects on close orbits. It is these objects that were favoured by the previous results, because they suggested that even a small mass at this location could still have a significant effect on the apparent outer companion elements.

We now examine the case where two observations are made at times $t=0$ and $\Delta t$. All bodies are again on coplanar circular orbits. We assume for simplicity that the time between observations is much smaller than the outer object period, which is valid as companions currently detectable by direct imaging generally exist far from their host star. This means that the motion of the outer object is approximately linear, with velocity $\mathbf{v'} \approx \Delta \mathbf{r'} / \Delta t$. Without loss of generality we can specify that at the time of the first observation the system is in the same configuration as for the single epoch case (i.e. the initial outer object position in the astrocentric frame is given by Equation 1 with $f = f_0$). At a time $\Delta t$ later, the observed companion will have a position

\begin{equation}
 \mathbf{r'}_{t=\Delta t} = a \left( \begin{array}{c}
      \cos (f_0 + \Delta f) \\
      \sin (f_0 + \Delta f)  \end{array} \right) + \mu a_{\rm i} \left(
\begin{array}{c}
      \cos(n_{\rm i} \Delta t) \\
      \sin(n_{\rm i} \Delta t)   \end{array} \right)
 \label{eq: case2 r'}
\end{equation}

\noindent in the new astrocentric frame, where $\Delta f = \sqrt{GM/a^3} \Delta t$ and $n_{\rm i} \equiv 2\pi/T_{\rm i}$. Therefore we may estimate $\mathbf{v'}$ as \mbox{$(\mathbf{r'}_{t=\Delta t}-\mathbf{r'}_{t=0}) / \Delta t$}, and using this and the companion's position in one of the images we may proceed in deriving its astrocentric elements as before. 

This time the elements are not only functions of $\mu$ and $\alpha$ but also of $\Delta t$, and the resultant solutions are more complicated than in the previous case. However to first order in $\mu$ (assuming $\alpha$ is small) these elements may be well approximated as 

\begin{equation}
 \frac{a'-a}{a} \approx \mu \left[1 + \frac{2}{\sqrt{\alpha}}\zeta(\Delta t)\cos \left(f_0 - \pi\frac{\Delta t}{T_{\rm i}} \right)\right]
\label{eq: case2 a' 1st order}
\end{equation}

\noindent and

\begin{align}
\begin{split}
& e'\approx \mu \left[1 + \frac{4}{\sqrt{\alpha}}\zeta(\Delta t)\cos\left(f_0-\pi\frac{\Delta t}{T_{\rm i}}\right)\right. \\
&\left.+ \frac{1}{\alpha} \zeta^2(\Delta t) \left(1 + 3 \cos^2\left(f_0- \pi\frac{\Delta t}{T_{\rm i}}\right)\right)\right]^\frac{1}{2},
\end{split}
\label{eq: case2 e' 1st order}
\end{align}

\noindent where

\begin{equation}
 \zeta(\Delta t) \equiv \sinc \left(\frac{\pi \Delta t}{T_{\rm i}} \right) .
\label{eq: zeta}
\end{equation}

\begin{figure}
  \centering
      \includegraphics[width=8cm]{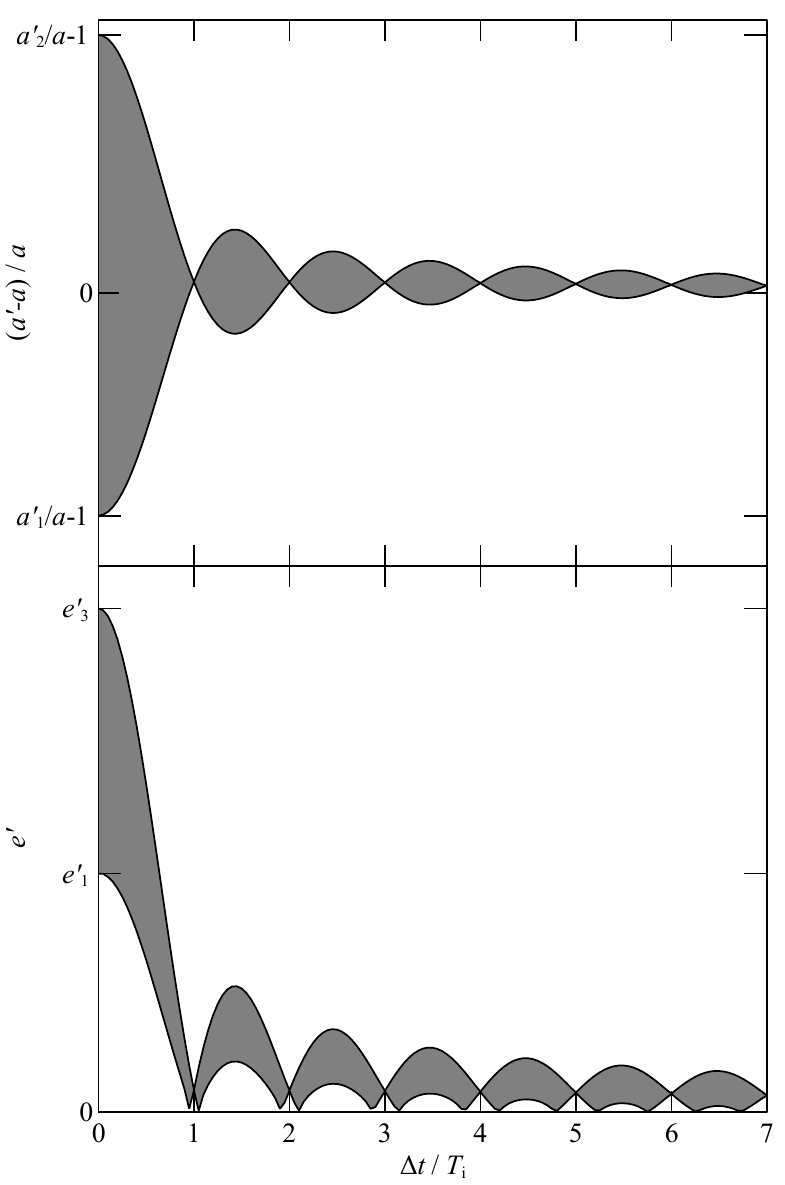}
  \caption{Osculating elements as functions of $\Delta t / T_{\rm i}$. For each value of $\Delta t / T_{\rm i}$ the astrocentric elements oscillate as functions of $f_0$ (similar to Figure \ref{fig: osculating elements}), and the range over which they oscillate is denoted here by the shaded region. This plot is system specific and has been produced using $\mu = 0.001$ and $\alpha = 0.008$, but is qualitatively the same for all parameters. The quantities on the vertical axes are the same as on Figure \ref{fig: osculating elements}; note that the range tends to that of the simpler case as $\Delta t \rightarrow 0$.}
  \label{fig: case2 elements vs dt}
\end{figure}

\noindent Note that in the limit $\Delta t \rightarrow 0$, $\zeta(\Delta t) \rightarrow 1$ and the above expressions reduce to Equations \ref{eq: a' 1st order} and \ref{eq: e' 1st order}. The behaviour of these elements as a function of $\Delta t$ (using the full calculation rather than the first order approximations above) is shown on Figure \ref{fig: case2 elements vs dt}.

There are several differences between this case and the $\Delta t/T_{\rm i} \approx 0$ regime described earlier. Firstly the phases of $(a'-a) / a$ and $e'$ are now shifted in $f_0$ when compared to the single epoch case, due to the changing object positions during the calculation. This manifests itself as the $f_0- \pi\Delta t/T_{\rm i}$ terms in the equations. Secondly the multiple epoch scenario reduces the amplitude of $(a'-a) / a$ and $e'$ when compared to the single epoch case; Figure \ref{fig: case2 elements vs dt} shows that the magnitude of these astrocentric elements show a long term decline as $\Delta t$ is increased. This can be explained by noting that the stellar motion, as well as the motion of the outer object, is effectively averaged by the use of multiple observation epochs. That is, as the observed astrocentric velocity is derived as $\mathbf{v'} = \Delta \mathbf{r'} / \Delta t$ where $\mathbf{r'} = \mathbf{r} - \mathbf{r_*}$, the apparent velocity shift caused by the stellar motion is therefore $\Delta \mathbf{r_*} / \Delta t$. For circular stellar motion the velocity derived in this way will always be smaller than the true velocity, and so the effect of this averaging is to reduce $\delta v$ and therefore the amplitude of the outer object's osculating elements. This manifests itself primarily as the $T_{\rm i}/\Delta t$ term in Equation \ref{eq: zeta}, which causes the long term $\sim 1/\Delta t$ declines in element amplitude visible on Figure \ref{fig: case2 elements vs dt}. Note that if $\Delta t / T_{\rm i} > 1$ the inner binary makes at least one complete revolution between observations, and therefore the apparent stellar velocity as ``seen'' by the outer object will be significantly reduced.

Figure \ref{fig: case2 elements vs dt} also shows that the elements undergo short term oscillatory behaviour as a function of $\Delta t$, and that the range over which they oscillate (and hence the dependence on initial outer object true anomaly, $f_0$) is zero when $\Delta t / T_{\rm i}$ is an integer. This is another effect of the apparently reduced stellar motion described above. Firstly when $\Delta t / T_{\rm i}$ is an integer, the star has the same position at both observation epochs, and so its apparent velocity is zero. As the fractional radial shift $\delta r / r$ caused by the unseen inner mass is negligible, in this case the outer object effectively ``sees'' the star with no velocity and very little offset from the barycentre, and so the apparent astrocentric elements do not depend on the initial true anomaly of the outer companion. The only difference between the two sets of elements is therefore caused by this small barycentric offset and the use of the star's mass to derive the astrocentric values, rather than the combined mass of the inner binary. Also when $\Delta t / T_{\rm i}$ is a half integer (apart from when $\Delta t / T_{\rm i} = 1/2$) the binary is observed to have advanced by half an orbit, and so $\Delta \mathbf{r_*}$ and hence $\delta v$ is maximum. This causes the sub maxima in $(a'-a) / a$ and $e'$ at these locations, although they may be slightly shifted due to the general $\sim 1/\Delta t$ decline. Note that for $\Delta t / T_{\rm i} < 1$ the maxima lies at $\Delta t / T_{\rm i} = 0$ rather than $1/2$; this is because the binary has not yet made one complete revolution and so $\Delta t / T_{\rm i} \rightarrow 0$ rather than a larger integer, and the apparent stellar velocity therefore tends to its true value. Finally there is a very slight downward trend in $(a'-a) / a$ as $\Delta t$ is increased (not very significant in Figure \ref{fig: case2 elements vs dt} but pronounced in some cases) caused by the breakdown of the linear motion approximation.

\begin{figure}
  \centering
      \includegraphics[width=8cm]{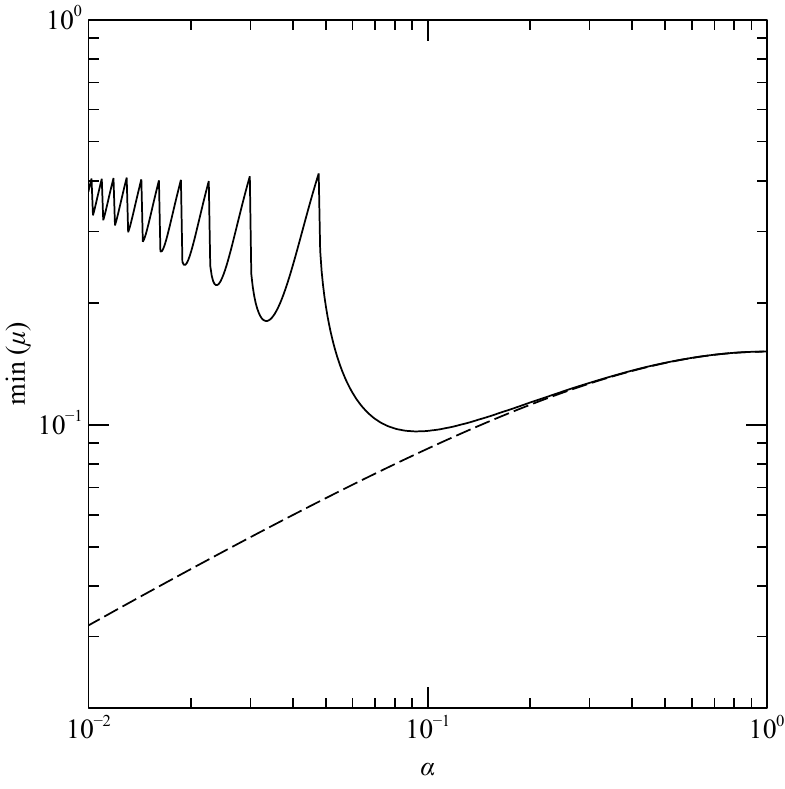}
  \caption{Minimum $\mu$ required to give an outer object on a circular orbit an apparent eccentricity of 0.8. The solid line is calculated numerically for the case where the observed companion's velocity is derived from two observations, and the dashed line shows the single observation regime. Note that this plot is quantitatively system specific and has been produced using $m_* = 1.92 {\rm M_\odot}$, $a = 120$ AU and $\Delta t = 7.6$ yrs.}
  \label{fig: case2 min mu}
\end{figure}

The important thing to note for this case is that any non-zero $\Delta t$ reduces the effect of an unseen inner mass when compared to the single epoch scenario, and that this effect becomes significant if this object lies close to the star. Figure \ref{fig: case2 min mu} shows the minimum $\mu$ as a function of $\alpha$ required to give the outer an astrocentric eccentricity as before, only now for an example set of parameters with $\Delta t \neq 0$. The specific masses and turning points are system dependent, however the plot is qualitatively the same for all parameters; at large $\alpha$, $\Delta t / T_{\rm i} \rightarrow 0$ and the result tends to the simpler regime of Section \ref{sec: case 1}. As $\alpha$ gets smaller the $\Delta t \neq 0$ case begins to dominate, and there is now a lower limit on $\mu$ to give a circular companion an apparent astrocentric eccentricity. Therefore $\mu$ may not be ever reduced simply by moving the inner object ever closer to the star. Finally the required mass increases sharply beyond this turning point, and sub-minima are also present due to $T_{\rm i}$ changing as a function of inner binary semi-major axis.

Proceeding as for the simpler case, we may estimate the minimum unseen mass as a function of $\alpha$ required to give an observed circular companion an apparent astrocentric eccentricity $e'$. By rearranging Equation \ref{eq: case2 e' 1st order} we derive an approximate expression for this minimum mass that is valid when $\mu$ and $\alpha$ are small:

\begin{equation}
 \mu \gtrsim e' \left(1 + \frac{4}{\sqrt{\alpha}} \zeta(\Delta t) + \frac{4}{\alpha} \zeta^2(\Delta t) \right)^{-\frac{1}{2}}.
\label{eq: case2 min mu 1st order}
\end{equation}

\noindent We may then differentiate this expression with respect to $\alpha$ and set it to zero to find the absolute minimum value of $\mu$. The first non-zero solution to the resulting equation occurs when

\begin{equation}
 \tan(\Gamma) = \frac{3}{2}\Gamma,
\label{eq: Gamma_min}
\end{equation}

\noindent where $\Gamma \equiv \pi \Delta t / T_{\rm i}$.

At this point as no information is known about the unseen inner mass it makes sense to remove the dependence on $T_{\rm i}$ from the above equations and replace it with $\tau$, the ratio of $\Delta t$ to the outer object period, which may be more intuitively estimated. Thus $\Delta t / T_{\rm i} = \tau /\alpha^{3/2}$ and $\Gamma = \pi \tau /\alpha^{3/2}$. Note that all of the system specific information (the star mass, $\Delta t$ and $a$) is contained within $\tau$. As a larger $\Gamma$ corresponds to a smaller value of $\alpha$, the smallest non-zero solution of Equation \ref{eq: Gamma_min} corresponds to the global minimum value of $\mu$. $\Gamma \approx 0.967$ at this point. Therefore an inner mass will have the greatest effect on the astrocentric elements of the outer if

\begin{equation}
\frac{\Delta t}{T_{\rm i}} \approx 0.31,
\label{eq: dt_Ti}
\end{equation}

\noindent i.e. the observational baseline is about a third of the inner object period. Substituting this into Equation 10, we find that in order for an unseen inner mass to give a circular outer object an astrocentric eccentricity $e'$

\begin{equation}
 \mu \gtrsim e' \left(1 + 2.30 \tau^{-1/3} + 1.32 \tau^{-2/3} \right)^{-\frac{1}{2}},
\label{eq: min_mu}
\end{equation}

\noindent and the location of this mass in order for $\mu$ to have the minimum possible value must be

\begin{equation}
 \alpha \approx 2.19 \tau^{2/3}.
\label{eq: min_a1_a}
\end{equation}

\noindent The latter equation is independent of $e'$, and so the radius at which the inner object has the greatest effect is only dependent on $a$ and $\tau$. The secondary minima on Figure \ref{fig: case2 min mu} correspond to higher $\Gamma$ solutions to equation \ref{eq: Gamma_min}, and the peaks correspond to a second set of $\alpha$-dependent roots to the differential of Equation \ref{eq: case2 min mu 1st order}. Whilst the above equations are only approximate, they agree well with minimum masses and corresponding semi-major axis ratios calculated numerically without any simplifications.

\begin{figure}
  \centering
      \includegraphics[width=8cm]{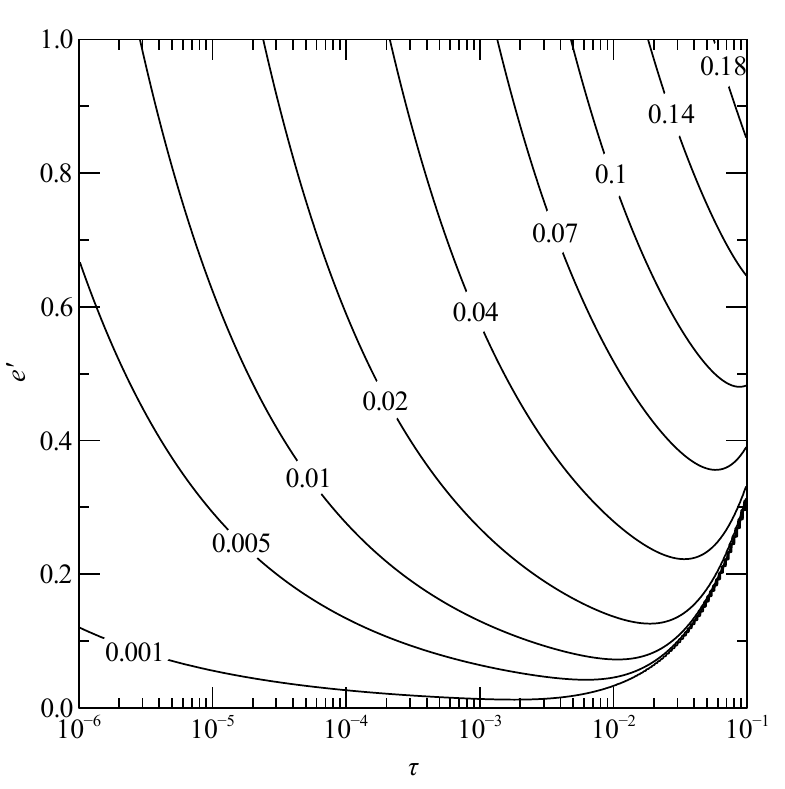}
  \caption{Minimum $\mu$ required to give an outer object on a circular orbit an apparent eccentricity $e'$, as a function of $\tau$. The graph was calculated numerically without making any approximations, but shows good agreement with Equation \ref{eq: min_mu} for $\tau \lesssim 10^{-2}$. Above this value the lines of arbitrary ${\rm min}(\mu)$ converge, as the error on the eccentricity caused by the assumption that the companion motion is linear becomes more significant than the error caused by an unseen mass.}
  \label{fig: case2 tau contours}
\end{figure}

Figure \ref{fig: case2 tau contours} shows the absolute minimum $\mu$ as a function of $\tau$ and $e'$ calculated by a numerical grid search, which shows good agreement with Equation \ref{eq: min_mu} for $\tau \lesssim 10^{-2}$. Above this value the two diverge as the linear motion approximation breaks down; assuming the companion moves in a straight line between epochs will always introduce an error on the derived elements, and this error will increase as a greater fraction of the orbit is observed. Therefore this effect is most apparent in the lower right corner of the plot, where the difference between the outer object's true velocity and that estimated linearly is sufficient to give the body an apparent eccentricity even in the absence of a third mass. This plot may be used to establish whether the apparent eccentricity of an observed companion could be entirely caused by the presence of an unseen inner mass.

The parameter $\tau$ is given by

\begin{equation}
 \tau = \frac{\Delta t}{2\pi}\sqrt{\frac{G(m_* + m_{\rm i})}{a^3}}
\label{eq: tau exact}
\end{equation}

\noindent and contains two unknowns, $m_i$ and $a$. However we again take advantage of $\delta r / r$ being small, and hence can make the approximation $r' \approx a$. Therefore $\tau$ may be accurately estimated as

\begin{equation}
 \tau \approx  \Delta t \sqrt{m_*} r'^{-3/2},
\label{eq: tau}
\end{equation}

\noindent where $m_*$, $\Delta t$ and $r'$ are in units of solar masses, years and AU respectively. This approximation may be used in all of the above calculations. As an example suppose two observations of an object at 100 AU from a solar type star are made 1 year apart, and orbital motion is detected between the epochs yielding an astrocentric eccentricity of 0.5. If the object is actually on a circular orbit then $\tau = 10^{-3}$, and thus from Figure \ref{fig: case2 tau contours} we see that ${\rm min}(\mu)$ is between 0.02 and 0.04 (the actual value is 0.035). Equation \ref{eq: min_a1_a} shows that in order for $\mu$ to have this minimum value, the inner body must be located at 2.2 AU.


\section{Outer object on elliptical orbit}
\label{sec: eccentric orbit}

\noindent We now generalise the above results to allow the outer object to have some eccentricity in the barycentric frame. As before an inner mass could potentially increase this eccentricity in the astrocentric frame. The apparent eccentricity may also now be decreased, i.e.  an unseen mass could also make the companion appear less eccentric than it actually is. However as circular orbits are generally favoured by planet formation models and highly eccentric companions point towards some disruptive dynamical event in the system's history, we only focus on increasing the companion's apparent eccentricity in this paper. The magnitude of this effect is expected to be roughly symmetrical, so an unseen mass could potentially increase or decrease the apparent eccentricity of an imaged companion by roughly the same amount. Therefore the size of the potential eccentricity underestimation may also estimated by the following method.

\subsection{Negligible time between observations}


\begin{figure}
  \centering
      \includegraphics[width=8cm]{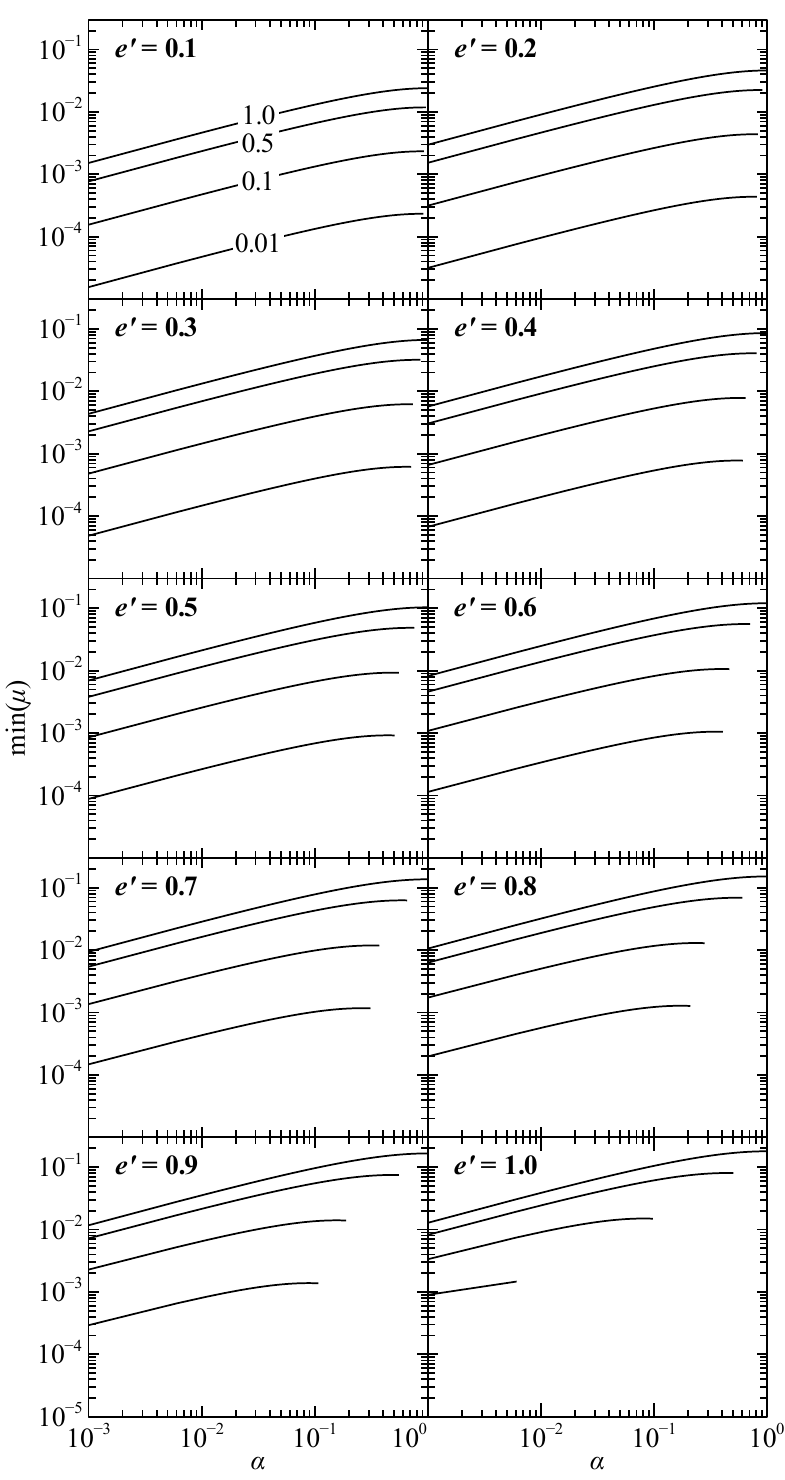}
  \caption{Minimum $\mu$ required to induce a given fractional error in observed eccentricity, $\Delta e/e' \equiv (e'-e)/e'$, as a function of $\alpha$ for different astrocentric eccentricities in the $\tau = 0$ regime. The lines show $\Delta e/e' = 0.01$, 0.1, 0.5 and 1.0 in the order shown on the $e' = 0.1$ panel, and may be estimated using the equation in the Appendix. The model is not valid if the outer object is inside the orbit of the inner at pericentre, so the values of $\alpha$ resulting in this scenario are omitted. For an observed $e'$, the reader may use this plot to determine the maximum error on the derived eccentricity given observational upper limits on $\mu$ as a function of $\alpha$.}
  \label{fig: general case 3}
\end{figure}

\noindent We will proceed as before, by first analysing the $\Delta t = 0$ regime and then extending this to the multiple epoch case. We again assume the orbits to be coplanar, and the inner binary orbit is still circular. Equations \ref{eq: r'} and \ref{eq: v'} now become

\begin{equation}
 \mathbf{r'} = \frac{a(1-e^2)}{1+e\cos(f)} \left( \begin{array}{c}
      \cos (\omega+f) \\
      \sin (\omega+f)   \end{array} \right) + \mu a_{\rm i} \left(
\begin{array}{c}
      1 \\
      0   \end{array} \right)
 \label{eq: r'_w}
\end{equation}

\noindent and

\begin{align}
\begin{split}
 \mathbf{v'} &= \sqrt{\frac{GM}{a(1-e^2)}} \left(
\begin{array}{c}
      - \sin (\omega + f) - e\sin \omega \\
      \cos (\omega + f) + e\cos \omega   \end{array} \right) \\
    &+ \mu \sqrt{\frac{GM}{a_{\rm i}}} \left( \begin{array}{c}
      0 \\
      1   \end{array} \right),
\end{split}
\label{eq: v'_w}
\end{align}

\noindent where $e$ is the barycentric eccentricity. As for the $e=0$ case, we use these equations to derive the outer object's astrocentric elements. The system now has a 2D phase, given by the argument of periapsis ($\omega$) and $f$. The change in elements is maximised when the difference between the stellar motion and that of the outer body is greatest, which occurs when the outer object is at pericentre. The maximum values of $a'$ and $e'$ therefore occur when $\omega = f = 0$. Proceeding as before we may again derive a lower limit on the unseen inner mass based on the observed companion's astrocentric eccentricity, which now depends on its assumed barycentric eccentricity.

Figure \ref{fig: general case 3} shows the maximum fractional error in observed eccentricity, $\Delta e/e' \equiv (e'-e)/e'$, as a function of $\mu$ and $\alpha$ for different observed astrocentric eccentricities. Note that if $\Delta e/e' = 1$ then the outer object is on a circular orbit, and also that the errors plotted are positive (i.e. $e' > e$). The contours were again calculated using the full expression and we also derive a simplified analytical expression equivalent to Equation \ref{eq: case2 min mu 1st order}, but it is cumbersome and so given in the appendix. Figure \ref{fig: general case 3} is analogous to Figure \ref{fig: case1 min mu} as they are both independent of star mass and semi-major axes.

The plot is qualitatively similar to Figure \ref{fig: case1 min mu}, as ${\rm min}(\mu)$ still follows a $\sqrt{\alpha}$ dependence and turns over as other $\alpha$ terms become non-negligible. We have neglected orbits for which the outer companion has pericentre interior to the orbit of the inner mass, as 3 body dynamics would also be important in this region and the results would be incorrect. As for Figure \ref{fig: case1 min mu}, this plot may be used to determine the maximum error on the derived eccentricity of a companion, given an observational upper limit on the inner mass as a function of orbital radius.


\begin{figure*}
  \centering
      \includegraphics[width=10.cm]{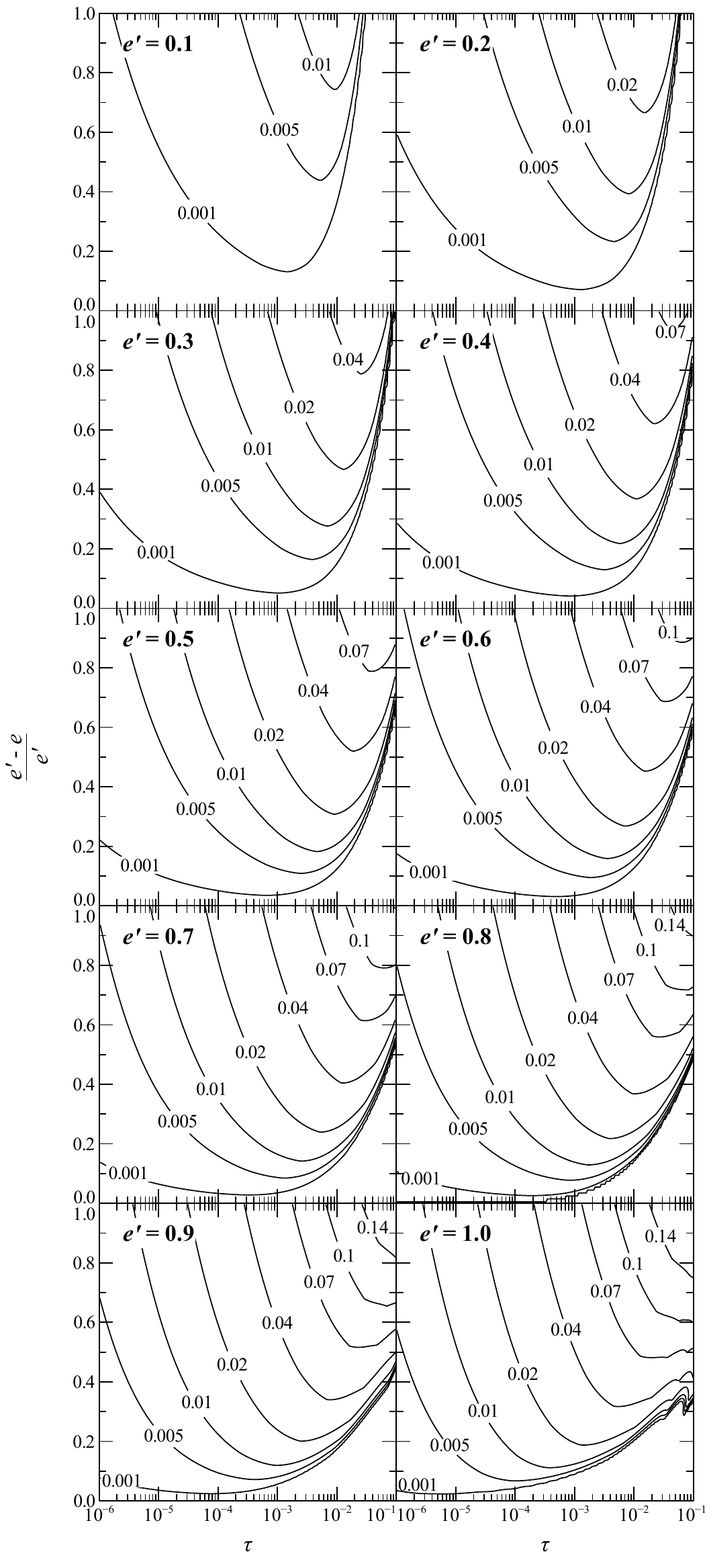}
  \caption{Minimum ${\rm min}(\mu)$ required to induce a given fractional error in observed eccentricity, $\Delta e/e'$, as a function of $\tau$ for different astrocentric eccentricities. The values of $\alpha$ corresponding to these minimum masses may still be found using Equation \ref{eq: min_a1_a}. Note the deviations from smooth contours on the right side of the $e' = 1.0$ plot are numerical in nature, and should be ignored.}
  \label{fig: de_vs_tau}
\end{figure*}


\subsection{Non negligible time between observations}

Once again, we consider the use of multiple observational epochs to derive the outer companion velocity. All analytics are now very inelegant and can be sensitive to simplifications, so there is little merit in reproducing them here. However the resulting plots of ${\rm min}(\mu)$ required to boost $e$ up to $e'$ as a function of $\alpha$ are qualitatively the same as Figure \ref{fig: case2 min mu}, but shifted down slightly as ${\rm min}(\mu)$ does not have to be as large. Figure \ref{fig: case2 min mu} is \textit{not} shifted in $\alpha$ by the introduction of non-zero $e$, i.e. the global minimum value of ${\rm min}(\mu)$ still occurs at the same ratio of semi-major axes as the $e=0$ case. This has been tested numerically across the entire parameter space. Therefore Equation \ref{eq: min_a1_a} may still be used locate the value of $\alpha$ where the inner mass will have the greatest effect, although Equation \ref{eq: min_mu} no longer holds.

As the introduction of multiple observations again leads to an absolute minimum value of $\mu$ required to give an outer object a given astrocentric eccentricity, we may produce a plot analogous to Figure \ref{fig: case2 tau contours} that shows this minimum mass as a function of $\tau$. This is presented on Figure \ref{fig: de_vs_tau} for various astrocentric eccentricities, found using a numerical grid search. Note that the behaviour for $\tau \gtrsim 10^{-2}$ is similar to that on Figure \ref{fig: case2 tau contours} due to the breakdown of the linear motion approximation.

This plot may be used to establish whether the apparent eccentricity of an observed companion could be incorrect due to the effect of an unseen inner mass. However there is one final problem; if the outer object may now have a barycentric eccentricity, we can no longer approximate the parameter $\tau$ in the same way as before because the barycentric semi-major axis is unknown. However we may constrain $\tau$ to lie along a line in $\Delta e / e'$ space for the best case scenario, so we can still find the minimum inner mass required to introduce a given error on the outer object eccentricity. As $\tau$ will be small for wide separation companions, the greatest change in orbital elements will occur if this body is near pericentre. In addition we know that $\delta r / r$ is small, so at this point $r' \approx a(1-e)$. Hence replacing $a$ with $[m_*(\Delta t/\tau)^2]^{1/3}$ will give the value of $\tau$ if the object is at pericentre, which is a function of $e$. Noting that $e = e'(1 - \Delta e / e')$ we can then rearrange this in terms of $\Delta e / e'$. Therefore if the outer object is at pericentre then
 
\begin{equation}
\frac{\Delta e}{e'} \approx 1 + \frac{r'}{e'} \left[\frac{1}{m_*}\left(\frac{\tau}{\Delta t}\right)^2 \right]^{\frac{1}{3}} -\frac{1}{e'} ,
\label{eq: de bound}
\end{equation}

\noindent where $m_*$, $\Delta t$ and $r'$ are again in units of solar masses, years and AU respectively. This equation may be over-plotted on the appropriate panel of Figure \ref{fig: de_vs_tau}, and hence the maximum eccentricity error an unseen mass may introduce will be at the $\Delta e / e'$ value where this mass contour crosses the above line.

\begin{figure}
  \centering
      \includegraphics[width=8cm]{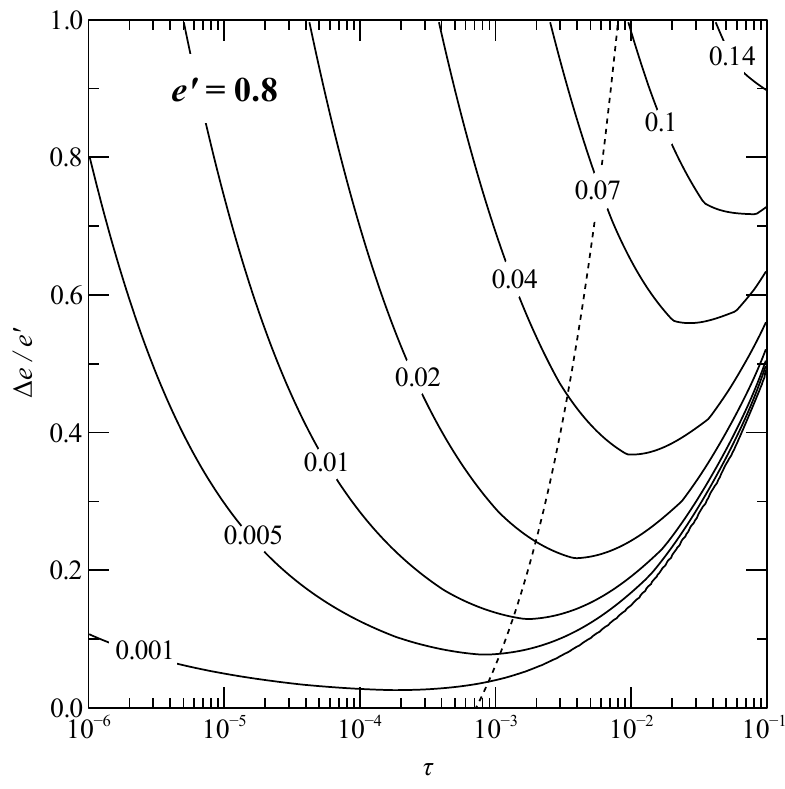}
  \caption{Example application of $\tau$ constraining using Equation \ref{eq: de bound}, with the parameters $m_*$, $\Delta t$, $r'$ and $e'$ equal to $1.92{\rm M}_\odot$, 7.6 yrs, 120 AU and 0.8 respectively. The maximum possible error in orbital elements for a given inner mass occurs where this contour crosses the dotted line (Equation \ref{eq: de bound}). For example if the maximum permitted $\mu$ is 0.04, then the maximum fractional error an inner mass could induce on the eccentricity would be 0.45.}
  \label{fig: with_tau_bounds}
\end{figure}

We plot line this for an example set of parameters on Figure \ref{fig: with_tau_bounds}, along with the appropriate panel of Figure \ref{fig: de_vs_tau}. For many systems, such as that plotted here, the possible range of $\tau$ will only span about an order of magnitude and so can be estimated fairly easily. A similar plot may be made by the user for their system of interest, and should be used to establish whether an unseen inner mass could introduce a significant error on the outer eccentricity. 

\section{Applicability}
\label{sec: applicability}

\noindent The scenario outlined in this paper, in which a directly imaged companion has incorrectly derived orbital elements due to the effect of an unseen inner mass on the stellar motion, will not be important for all directly imaged systems. In this section we examine what criteria a system must fulfil in order for this scenario to warrant consideration.

We require the system to harbour a directly imaged companion for which orbital motion has been detected, which is also much less massive than its parent star. We also require the parameter $\tau$ to be small in order for the inner body to have the greatest effect; Figures \ref{fig: case2 tau contours} and \ref{fig: de_vs_tau} show that $\tau$ must also be of order $10^{-2}$ or smaller if the primary effect on $e'$ is to be caused by an inner mass rather than a breakdown of the linear motion assumption. However there are bounds on the minimum value of $\tau$ for a given system, set by observational limitations. Firstly $\Delta t$ may not take any arbitrary value; in reality we have a maximum observational baseline, $\max(\Delta t)$. Substituting this into the equation for $\tau$, we arrive at an upper limit of 

\begin{equation}
\tau < \max(\Delta t) \sqrt{\frac{m_*}{a^3}}
\label{eq: tau upper bound}
\end{equation}

\noindent where $\max(\Delta t)$, $m_*$ and $a$ are in units of years, solar masses and AU respectively. There is also a lower limit on $\tau$, which arises because the difference in the angular separation of the companion between the two observational epochs must be large enough to be resolvable. If the orbit of this object is eccentric, then the largest change in angular position will occur if the orbit is face on with the object at pericentre. If $\tau$ is small, we may approximate the change in true anomaly to be $\Delta f \approx 2\pi \tau \sqrt{(1+e)/(1-e)^3}$ at this point. Between the two epochs the companion will move by an angular distance of approximately $[a(1-e)/d] \Delta f$ at pericentre as viewed from Earth, where $d$ is the distance to the system. Therefore the lower bound on $\tau$ is given by

\begin{equation}
\tau >  \frac{3\sqrt{2} \theta_{\rm cen}}{2\pi}\sqrt{\frac{1-e}{1+e}}\frac{d}{a},
\label{eq: tau lower bound}
\end{equation}

\noindent where $\theta_{\rm cen}$ is the $1\sigma$ half width centroiding accuracy. The factor of $3\sqrt{2}$ comes from the requirement of a three sigma detection of orbital motion. If $\theta_{\rm cen}$ is in radians then $d$ and $a$ must be in the same units; alternatively if $\theta_{\rm cen}$ is in arcseconds then $d$ and $a$ are in parsecs and AU respectively.

The above equations show that for this scenario to be potentially important the observed system must be nearby (small $d$) with a wide separation companion, but not so wide that orbital motion is undetectable. Eliminating $\tau$ from the above equations, for orbital motion to be detected the semi-major axis must fulfil

 \begin{equation}
a < m_* \frac{1+e}{1-e} \left(\frac{2\pi \max(\Delta t)}{3\sqrt{2} \theta_{\rm cen} d} \right)^2 
\label{eq: a max}
\end{equation}

\noindent (again, $3\sqrt{2}$ comes from the requirement of a three sigma detection of orbital motion).

Additionally an absolute lower limit on $a$ is $d \theta_{\rm res} / 2$, where $\theta_{\rm res}$ is the full width instrument angular resolution. This arises from the observable star-companion separation, and is not $m_*$ or $e$ dependent. This is a lower bound because in general the detection of companions close to the star is contrast limited as opposed to resolution limited. Therefore only very massive companions may be observed down to the ``currently unresolvable" limit, and lower mass objects must lie farther out to be detected. For example whilst the resolution of the HST is 0.2'', its effective inner working angle is actually about 0.7'' in the infrared \citep{Krist06}. The dependence of detectability on mass is not an issue for the outermost companions in this paper, because the analysis presented here is independent of this quantity provided that the imaged object is not massive enough to significantly perturb the inner binary. However this dependence will affect our ability to detect an inner object, as even a significant mass may be lost in the glare of the star if its semi-major axis is small enough (see Section \ref{sec: detectability}).

\begin{figure*}
  \centering
      \includegraphics[width=12cm]{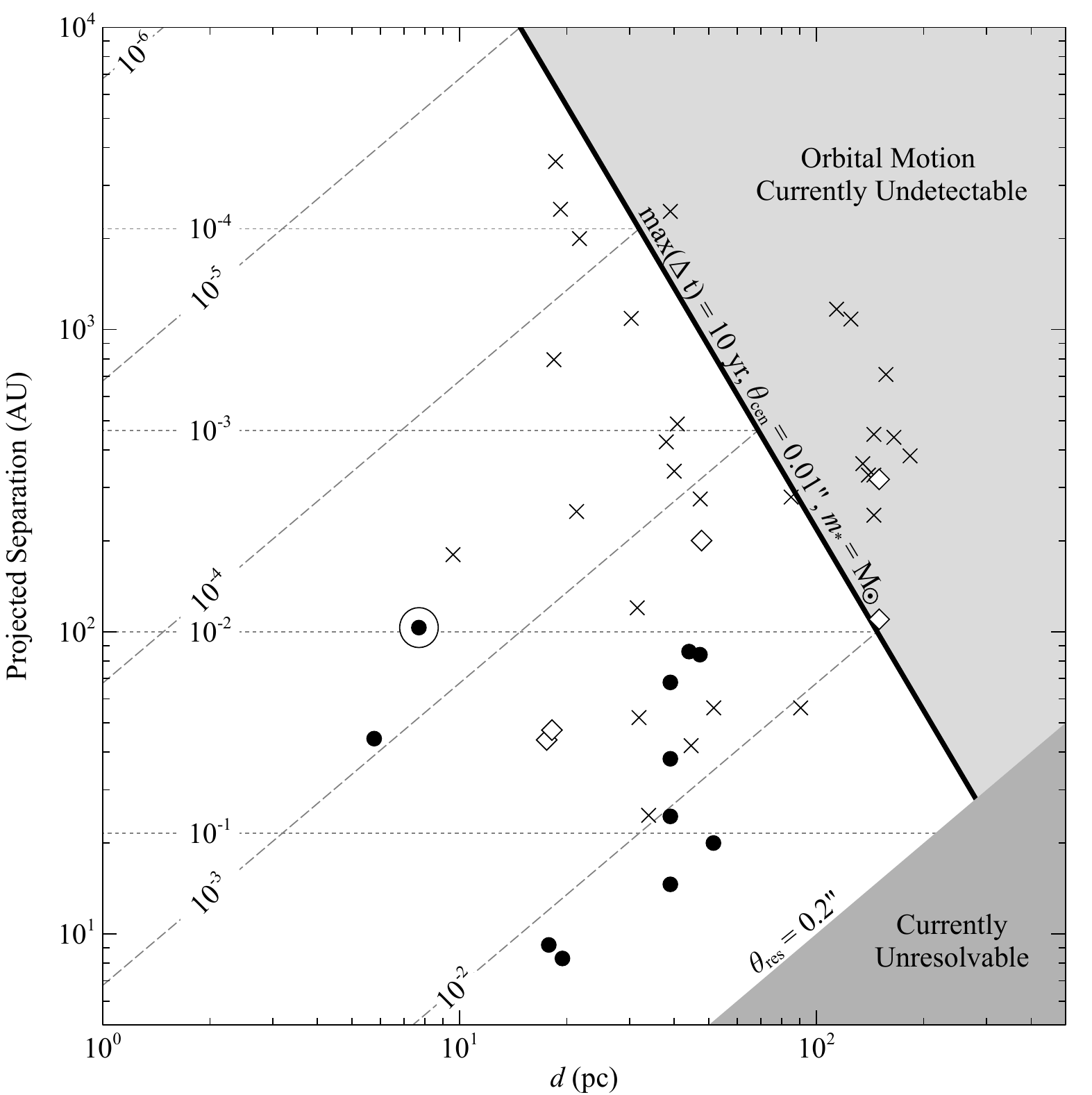}
  \caption{Projected separations of directly imaged sub-stellar companions. Circles and diamonds denote objects with probable ($\geq 3 \sigma$) and possible detections of orbital motion respectively. Crosses show companions with measured orbital motions consistent with zero. The two shaded regions show the areas where companions are currently unresolvable and where orbital motion is usually too small to be detected, both derived using parameters typical of modern observations. The two companions lying in the latter region with possible detections of orbital motion have unusually good centroiding accuracies of $\sim$ 1 mas, and motion has not been detected to $3 \sigma$. Note that the minimum separation required for detection will be higher than the limit from resolution alone due to contrast effects, but not by enough to affect the conclusions of this work. The diagonal dashed and horizontal dotted lines show the minimum and maximum values of $\tau$ respectively, for $1 {\rm M_\odot}$ stars using a 10 yr baseline (Equations \ref{eq: tau upper bound} and \ref{eq: tau lower bound}). For example, if Fomalhaut-b (circled) were on a circular orbit, $\tau$ would lie between $10^{-2}$ and $10^{-4}$. A lower $\tau$ indicates a higher susceptibility to the scenario described in this paper (see Figures \ref{fig: case2 tau contours} and \ref{fig: de_vs_tau}).}
  \label{fig: a_d_tau_plot}
\end{figure*}

Figure \ref{fig: a_d_tau_plot} shows all of the above limits on $a$ as functions of $d$ for the $e=0$ case, as well as the maximum and minimum values of $\tau$ from Equations \ref{eq: tau upper bound} and \ref{eq: tau lower bound}. We assume a maximum baseline of 10 yrs and a resolution of $0.2''$, that of the HST, for the above equations. The centroiding accuracy is taken to be $0.01''$, which may be reached by current observations (e.g. \citealt{Golimowski98, Kasper07, Neuhauser08, Neuhauser10}). In fact some observations have achieved even better accuracies, however we will use $0.01''$ as a typical value because we feel it is a better representation of the current level of precision. We also plot the projected separations and distances for a selection of sub-stellar companions detected by imaging (compiled from \citealt{Reid01, Wilson01, Metchev04, Metchev06, Chauvin05b, Zuckerman09, Tanner10, Rodriguez12} and \citealt{exoplanet.eu}). Companions with probable ($\geq 3\sigma$) and possible detections of orbital motion are highlighted, and we give details of these objects in Table \ref{tab: orbital_motion_companions}. This should give the reader a feel for the region of parameter space occupied by these objects, in relation to that which may be important for the scenario described in this paper. Note that we have not plotted the astrocentric semi-major axes of the companions; we assume nothing about their orbits or orientations. However the projected separations should provide order of magnitude approximations of $a$ sufficient for this plot.

\begin{center}
\begin{table*}
\begin{tabular}{c c c c}
\hline
Companion & $d$ (pc) & Projected Separation (AU) & Reference \\
\hline
\multicolumn{4}{c}{Probable ($\geq 3 \sigma$) detection of orbital motion} \\											
2M 0103(AB)-b	&	47	$\pm$	3	&	84	$\pm$	5$^{\rm a}$	&	\cite{Delorme13}	\\
$\beta$ Pic-b	&	19.44	$\pm$	0.05	&	8.3	$\pm$	0.3	&	\cite{Chauvin12}	\\
Fomalhaut-b	&	7.70	$\pm$	0.03	&	103.2	$\pm$	0.5	&	\cite{Kalas13}	\\
Gl 229-B	&	5.77	$\pm$	0.04	&	44.3	$\pm$	0.3	&	\cite{Golimowski98}	\\
HR 7672-B	&	17.8	$\pm$	0.1	&	9.2	$\pm$	0.1	&	\cite{Crepp12}	\\
HR 8799-b	&	39	$\pm$	1	&	68	$\pm$	2	&	\cite{Marois08}	\\
-c	&	-			&	38.0	$\pm$	1.0	&	-	\\
-d	&	-			&	24.5	$\pm$	0.6	&	-	\\
-e	&	-			&	14.6	$\pm$	0.4	&	\cite{Marois10}	\\
PZ Tel-B	&	52	$\pm$	3	&	20	$\pm$	1	&	\cite{Mugrauer12}	\\
TWA 5-B	&	44	$\pm$	4	&	86	$\pm$	2$^{\rm a}$	&	\cite{Neuhauser10}	\\
											
\hline											
\multicolumn{4}{c}{Possible ($< 3 \sigma$) detection of orbital motion} \\											
$\eta$ Tel-B	&	48	$\pm$	2	&	200	$\pm$	6	&	\cite{Neuhauser11}	\\
GJ 504-b	&	17.56	$\pm$	0.08	&	43.9	$\pm$	0.5	&	\cite{Kuzuhara13}	\\
GQ Lup-B	&	150	$\pm$	50	&	110	$\pm$	40	&	\cite{Neuhauser08}	\\
GSC 06214−00210-b	&	150	$\pm$	10	&	320	$\pm$	30	&	\cite{Ireland11}	\\
HD 130948-B, -C	&	18.2	$\pm$	0.1	&	47.3	$\pm$	0.3$^{\rm b}$	&	\cite{Ginski13}	\\
\hline

\end{tabular}

\caption{Distances and projected separations for a selection of imaged sub-stellar companions for which orbital motion may have been detected. $^{\rm a}$Central object is a multi-star system, and companion separation is given from the system barycentre. $^{\rm b}$Companion is a binary, and separation is given from the central mass to the binary barycentre. }
\label{tab: orbital_motion_companions}
\end{table*}
\end{center}

As already stated, the effect of an inner mass on the derived elements of an outer companion will be most important if $\tau$ is small. This condition means that systems most susceptible to this effect would lie as high up Figure \ref{fig: a_d_tau_plot} as possible (but below the upper limit for detectable orbital motion). Figures \ref{fig: case2 tau contours} and \ref{fig: de_vs_tau} suggest that $\tau \lesssim 10^{-2}$ for the inner body to significantly affect the derived elements of the outer body. It is clear that the area of parameter space where this scenario could be applicable is well populated by companions, so could be important for around half of currently imaged systems. Also note that the solid line scales as ($\max(\Delta t)/\theta_{\rm cen})^2$, so as observational techniques improve and achieve longer time baselines, more objects could be discovered that would be susceptible to this scenario.

The next generation instruments GPI and SPHERE (\citealt{Macintosh06} and \citealt{Dohlen06} respectively) are expected to discover many companions within 100 AU of young stars at 30 - 50 pc. Whilst not populating the upper regions of Figure \ref{fig: a_d_tau_plot}, many of these objects could still be susceptible to eccentricity errors caused by unseen inner masses. Furthermore as better centroiding precisions are achieved by these projects and others, orbital motion will be detectable using observations covering smaller fractions of companion orbits. This means that lower $\tau$ values will be reached, and hence these objects would be more susceptible to eccentricity errors induced by unseen inner masses. Therefore we conclude that the effect described in this paper could already be significant for many directly imaged systems, and will be applicable to more imaged companions in the future.


\section{How to use this paper}
\label{sec: how to use}

\subsection{Suggested Method}

\noindent In the above three sections we described the set-up of the problem. We now suggest a step-by-step method to establish whether a derived astrocentric eccentricity is likely to be incorrect due to the presence of an inner object. We then apply this method to some example systems.

\begin{enumerate}
 \item Is the system suitable? It should have a known stellar mass and at least two images of the companion, between which orbital motion is observed. The user should also have estimates of the de-projected companion-star separation $r'$ and astrocentric eccentricity $e'$. If the inclination is unknown then a lower bound on this eccentricity may be derived by varying the assumed line of sight position and velocity until a minimum is found. If the stellar rotation axis is known then the inclination may be estimated by assuming that the star and companion are coplanar (e.g. \citealt{LeBouquin09, Watson11, Kennedy13}). The system should also lie above the horizontal $\tau < 10^{-1}$ line on Figure \ref{fig: a_d_tau_plot}.

\item Could the companion be on a circular barycentric orbit? To establish this, estimate $\tau$ using Equation \ref{eq: tau} and find the minimum inner mass required to give the observed $e'$ using Figure \ref{fig: case2 tau contours}. Calculate the semi-major axis of this mass using Equation \ref{eq: min_a1_a}; if an object with this mass and location cannot be ruled by observation then it is possible for the imaged object to be on a circular orbit about the star-inner mass barycentre. If the inner mass is observationally excluded in this region but a larger mass may exist further out, use Figure \ref{fig: case1 min mu} or Equation \ref{eq: case1 min mu} to see if this object may lie farther from the star. Remember that these relations are not valid all the way up to $\alpha = 1$ due to three body dynamics. Also note that the mass may exist closer in than the Equation \ref{eq: min_a1_a} value, but as the required mass rises steeply with decreasing distance (Figure \ref{fig: case2 min mu}) this is a less useful consideration.

\item If the companion must be eccentric, what is the maximum error in this barycentric eccentricity that could be caused by an inner body? Firstly estimate $\tau$ as a function of $\Delta e/e'$ using Equation \ref{eq: de bound}, similar to that plotted on Figure \ref{fig: with_tau_bounds}. Locate the graph on Figure \ref{fig: de_vs_tau} that corresponds to the observed value of $e'$, and overlay this $\tau$ constraint onto it. For each combination of ${\rm min}(\mu)$ and $\tau$ that lies in along this line establish whether such an inner mass could exist at a distance given by Equation \ref{eq: min_a1_a}, noting that $a$ will have to be estimated as $r' / (1-e)$. If it may, then use Figure \ref{fig: de_vs_tau} to read off the maximum eccentricity error $\Delta e / e'$ corresponding to this combination of $\tau$ and ${\rm min}(\mu)$. Again, if the inner object cannot exist in this region but a larger mass cannot be excluded further out, use Figure \ref{fig: general case 3} or the equation in the appendix to calculate this minimum mass.

\end{enumerate}

If an observer wishes to minimise the effect of this scenario on a system, the best technique would be make multiple observations over a long baseline. This would increase $\tau$ and thus reduce the average stellar velocity. The use of more than two observations would also reduce the effect of the third body, as the observed astrocentric elements would oscillate over time and so would vary depending on the pair of observations used to derive them. This could be used in some cases to exclude inner masses with periods shorter than the longest baseline.

\subsection{An example: the Fomalhaut system}

\noindent Figure \ref{fig: a_d_tau_plot} shows that the area of separation-distance parameter space where an unseen inner mass could affect the derived eccentricity of an observed object is well populated by companions. However orbital motion has not been detected for many of these objects as the required observations have yet to be made. If additional measurements are taken in the near future, for example in an attempt to build up an exoplanet/brown dwarf eccentricity distribution, then the presence of unseen masses could induce significant errors on this distribution. However until such measurements are made the most susceptible planetary system is Fomalhaut, which we will use here as an example to demonstrate the above method.

The star has a mass of $1.92 {\rm M}_\odot$, with a directly imaged planet (Fomalhaut-b) at 100 AU in projection for which orbital motion has been observed over four epochs between 2004 and 2012 \citep{Kalas13}. $\tau$ therefore lies between $10^{-4}$ and $10^{-2}$ from Figure \ref{fig: a_d_tau_plot}. The system also contains a narrow debris disk \citep{Kalas05}; if the planet is assumed to lie in the plane of this disk then its astrocentric position and velocity are $\sim 120$ AU and $\sim 1$ AU/yr respectively, yielding an astrocentric eccentricity of about 0.8 \citep{Kalas13}. Thus this system is suitable for the method outlined in this work.

We first test the hypothesis that the planet's orbit is actually circular in a barycentric frame and is aligned with the disk. Using Equation \ref{eq: tau} with $\Delta t = 7.6$ yrs we calculate $\tau$ to be 0.008. Figure \ref{fig: case2 tau contours} shows that for Fomalhaut-b to be on a circular orbit with an inner mass giving it an astrocentric eccentricity of 0.8, the unseen inner companion must have $0.07 < {\rm min}(\mu) < 0.1$ for this value of $\tau$. Equation \ref{eq: min_a1_a} shows that such a planet would exist at $\alpha = 0.09$, and therefore $a_{\rm i} = 11$ AU. Such a planet is ruled out by photometric non-detections \citep{Kenworthy09,Kenworthy13}, which place a model dependent upper mass limit for $a_{\rm i} > 5$ AU of $12-20 {\rm M_J}$ ($\mu \sim 0.006$ -- 0.01). This limit is an order of magnitude lower than required, so Fomalhaut-b cannot be on a circular orbit coplanar with the disk with its apparent eccentricity caused by an inner planet.

We now relax the condition that Fomalhaut-b must lie in the disk plane, to establish whether it is possible for the companion to have a circular orbit in any orientation. By varying the assumed line of sight position and velocity components of the companion we derive a lower bound on its astrocentric eccentricity, which is is 0.5--0.8 depending on the pair of observation epochs used. If we assume the orbital plane that gives $e' = 0.5$, then $a = 166$ AU and $\Delta t = 1.7$. Therefore $\tau = 0.001$, and Figure \ref{fig: case2 tau contours} shows us that in order for Fomalhaut-b to be circular with this apparent eccentricity ${\rm min}(\mu)$ must be between $0.02$ and $0.04$. This is still higher than the upper bound from observations, and so Fomalhaut-b has a barycentric eccentricity regardless of the chosen orbital plane.

Given that the orbit of Fomalhaut-b cannot be circular, we now wish to establish the maximum error in its barycentric eccentricity that could be caused by an observationally allowed unseen mass. We revert to the case where the planet and disk are coplanar. We first use Equation \ref{eq: de bound} to constrain $\tau$ as a function of barycentric eccentricity, as we may no longer estimate $\tau$ using Equation \ref{eq: tau} because we have no information about the true semi-major axis. Note that Equation \ref{eq: de bound} again assumes the companion to be at pericentre as this is the most favourable case for the scenario described in this paper. We overlay this $\tau$ constraint on the $e' = 0.8$ plot from Figure \ref{fig: de_vs_tau}, which is shown for the relevant parameters on Figure \ref{fig: with_tau_bounds}. We know from the observational upper limits that $\mu \leq 0.006$, so the maximum value of $\Delta e / e'$ that this mass may induce occurs when the $\tau$ constraint intersects the $\mu = 0.006$ contour. This is at $\Delta e / e' = 0.088$, which corresponds to $e \geq 0.73$. Therefore using Equation \ref{eq: dt_Ti} we see that a $12 {\rm M_J}$ mass at $a_{\rm i} = 10$ AU could introduce a $10\%$ error on Fomalhaut-b's eccentricity.

Note that as we have four observations of Fomalhaut-b, we know that the linear velocity approximation is still good over at least eight years. Even though the inner planet / brown dwarf described above would have a period of about twenty years, we cannot use b's constant velocity to rule out such an object because the changing velocity of the star itself would be undetectable; the star would have moved by about $2\mu \alpha a = 0.1$ AU ($0.01''$ at 7.7 pc) over eight years, which would change the planet's velocity by $1\%$ and would therefore be undetectable with the current precision. Furthermore the uncertainties in $e'$ from the astrometry and assumptions about the orbital plane are large enough that the use of more than one pair of observations cannot rule out the scenario outlined above. We therefore conclude that, whilst Fomalhaut-b cannot be on a circular barycentric orbit, an unseen 12 ${\rm M_J}$ companion at 10 AU could result in a $\sim 10\%$ overestimation of its astrocentric eccentricity, so it serves as a good example of a possible use of the above method.

\subsection{Other example systems}

In addition to Fomalhaut-b, we also applied the method to other applicable systems from Table \ref{tab: orbital_motion_companions}. We excluded $\beta$ Pic-b as its $\tau$ value is too high, and HR7672-B due to RV constraints on the inner mass. We also excluded the 2M 0103(AB), HR 8799 and TWA 5 systems from analysis as they are known to host more than one companion, and the current method is therefore unsuitable. Note however that we do include HD 130948, in which the companion itself is a known binary, by treating the pair as a single object.

All of these companions have orbital motion which appears linear over the observational baseline, and their sky plane positions and velocities were therefore derived by fitting linear trends. However unlike Fomalhaut, the majority of these systems do not have any additional information with which to predict the plane of the companion's orbit. $\eta$ Tel does have a debris disk which is close to face on, so if the companion / disk are aligned then the former may be assumed to orbit in the sky plane \citep{Smith09}. However it is not clear that the two objects should necessarily be aligned, as the companion lies much farther from the star than the disk. Therefore we derive astrocentric eccentricities for all companions in two extreme cases. In the first case we assume that the orbit is constrained to the sky plane, and in the second case we assume that the orbit is orientated in such a way that the astrocentric eccentricity is minimised. Assuming the companions are actually on circular orbits, we calculate the locations and masses of the lightest inner objects required to give the observed eccentricities. The results are shown in Table \ref{tab: other_example_systems}.

\begin{center}
\begin{table*}
\begin{tabular}{c c c c c c c c c c c}

\hline																					
\multirow{2}{*}{Companion}	&	\multicolumn{5}{c}{Face on orbit}									&	\multicolumn{5}{c}{Minimum astrocentric eccentricity orbit}									\\
	&	$e'_{I = 0}$	&	$\tau$	&	$m_{\rm i}$ (${\rm M_J}$)	&	$a_{\rm i}$ (AU)	&	Allowed?	&	$e'_{\rm min}$	&	$\tau$	&	$m_{\rm i}$ (${\rm M_J}$)	&	$a_{\rm i}$ (AU)	&	Allowed?	\\
\hline																					
Gl 229-B	&	0.9	&	0.003	&	47	&	1.8	&	Yes	&	0.3	&	0.002	&	15	&	1.8	&	Yes	\\
PZ Tel-B	&	1.0	&	0.04	&	210	&	5.5	&	-	&	0.6	&	0.06	&	130	&	5.5	&	-	\\
$\eta$ Tel-B	&	1.0	&	0.006	&	260	&	14	&	No	&	0.0	&	0.006	&	-	&	-	&	-	\\
GJ 504-b	&	0.3	&	0.004	&	49	&	2.6	&	Yes	&	0.0	&	0.004	&	-	&	-	&	-	\\
GQ Lupi-B	&	0.9	&	0.002	&	50	&	3.7	&	Yes	&	0.9	&	0.002	&	50	&	3.7	&	Yes	\\
GSC 06214-00210-b	&	0.2	&	0.0004	&	5	&	3.7	&	Yes	&	0.0	&	0.0003	&	-	&	-	&	-	\\
HD 130948-B, -C	&	1.0	&	0.03	&	18	&	10	&	No	&	0.0	&	0.030	&	-	&	-	&	-	\\
\hline

\end{tabular}

\caption{Locations and masses of the least massive inner objects required to give observed companions certain astrocentric eccentricities, if the companions are in fact on circular orbits. The eccentricities in the $e'_{I = 0}$ column have been derived with their orbits confined to the sky plane, and those in the $e'_{\rm min}$ column have been calculated by varying the assumed line of sight position and velocity until a minimum $e'$ was found. The ``Allowed?'' column states whether this inner mass is observationally permitted. Note that $\eta$ Tel has a debris disk which lies roughly in the sky plane. We found no upper mass limits in the literature for companions $\sim 5$ AU from PZ Tel-A.}
\label{tab: other_example_systems}
\end{table*}
\end{center}

Firstly we consider the case where the orbits are assumed to lie in the sky plane. Four of these companions (Gl 229-B, GJ 504-b, GQ Lupi-B and GSC 06214-00210-b), in spite of their high astrocentric eccentricities, could actually be on circular orbits with eccentricity errors introduced by observationally allowed inner masses. Furthermore, whilst the inner masses required for $\eta$ Tel-B and HD 130948-B, -C to be circular in the sky plane are observationally excluded, the maximum allowed inner masses could introduce eccentricity errors of $30\%$ and $20\%$ respectively. We found no upper mass limits in the literature for companions $\sim 5$ AU from PZ Tel-A, however the $\sim$ 200 ${\rm M_J}$ inner mass required to cause a large eccentricity error should be easily detectable. Therefore this scenario may be quickly confirmed or excluded for this system.

In the second case, where the assumed line of sight position and velocity are varied, we see that many of the companions could actually be on circular orbits. However three of the systems must have some astrocentric eccentricity regardless of orbital plane. Of these, Gl 229-B and GQ Lupi-B could be on circular orbits with their apparent eccentricity induced by unseen inner masses. Again we have no upper mass limits from the literature for companions close to PZ Tel-A, however the 130 ${\rm M_J}$ required could well be detectable with current instruments.

It is clear that the inner objects required for this scenario are typically tens of Jupiter masses, and many of them would inhabit the brown dwarf desert which could make their existence unlikely \citep{Marcy00}. However it must be noted that brown dwarfs are occasionally observed in these locations (e.g. \citealt{De Lee13}), and hence such objects may not be rejected purely due to this consideration.


\section{Detectability of the unseen mass}
\label{sec: detectability}

\noindent Throughout this paper we have required a massive, unseen inner object to significantly affect the orbital elements of an outer companion. Such large masses may often be ruled out using observational constraints, however we emphasise that this is by no means the case for all systems. We now summarise several detection methods which may provide upper limits on these masses. 

Firstly there are limits from the images themselves. Whilst direct imaging is the best means to detect wide separation companions, it is less suited to objects closer to the star owing to the huge contrast between the stellar flux and that of a smaller mass. The contrast ratio between a Jupiter mass planet and a solar type star is $10^{-7}$ in the infrared \citep{Traub10}, and that of a brown dwarf to such a star is $10^{-3}$ to $10^{-6}$ for L0 to T6 dwarfs respectively (Figure 2.9, \citealt{Bernat12}). Whilst such contrast sensitivities are just beginning to be reached at wide angles from stars, companion detectability rapidly worsens closer in. For example the Gemini Deep Planet Survey of young, nearby FGKM stars did achieve contrast sensitivities of $10^{-7}$ in some cases, but this value degraded sharply within $4''$ of the stars to around $10^{-5}$ at $1''$ \citep{Lafreniere07}. Similarly the International Deep Planet Survey of A and F stars, which also focussed on young nearby systems, reached contrast ratios of $10^{-7}$ for some objects but only at separations greater than $6''$ \citep{Vigan12}. At the distances of stars with wide separation imaged companions (Figure \ref{fig: a_d_tau_plot}) these angular scales correspond to tens or hundreds of AU, which are much larger than the semi-major axes of masses typically required to introduce a significant error on an outer body's eccentricity. 

Projects such as SPHERE \citep{Dohlen06}, GPI \citep{Macintosh06} and Project 1640 \citep{Hinkley08} should significantly increase the contrast sensitivity close to the star, and could rule out some of the inner companions required in this paper. However even these instruments would struggle to identify objects of several to tens of Jupiter masses 5--10 AU from stars at 50 pc (Figure 4, \citealt{Beichman10}). A further problem lies in actually converting these contrast sensitivities into upper mass limits; this is not a problem in older systems, but is a challenge for objects orbiting young stars. This is because at early ages ($\lesssim 100$ Myr) hot and cold start models produce significantly different estimates of a companion's luminosity \citep{Spiegel12}. Young companions are also brighter \citep{Marley07} and hence more likely to be detected, which means that this problem is commonly encountered. All these considerations mean that for now direct imaging is not the best means to locate or exclude inner objects, although in many cases it provides the only mass constraint in the absence of any other detection method being applied. 

The best upper limits on the masses of potential inner companions are likely to come from the radial velocity (RV) technique, which is very good at detecting large objects orbiting close to the star. For a circular inner mass to be undetected to $3\sigma$ if RV data is available for at least half an orbit,

\begin{equation}
\mu \sin i \lesssim 3 \times 10^{-4} \left(\frac{m_*}{\rm 1 M_\odot}\right)^{-1/2}\left({\frac{a_{\rm i}}{\rm 10 AU}}\right)^{1/2} \frac{K}{1 \rm m/s}
\label{eq: RV_detection_limit_simple}
\end{equation}

\noindent where $i$ is the inclination ($i=0$ being face on) and $K$ is the $1\sigma$ radial velocity sensitivity. $K$ is of the order of 1 m/s for current techniques \citep{Pepe11}. Whilst this method is therefore sensitive enough to rule out many unseen masses of the type described in this paper it is not without its limitations. Most importantly, the detection limit in Equation \ref{eq: RV_detection_limit_simple} becomes significantly degraded when the observational baseline is longer than the inner object's orbital period. Indeed, even if the acceleration from the companion remains at a detectable level, the interpretation of such long-term RV trends remains unknown until one full orbital period has been sampled (e.g. \citealt{Crepp13}). In addition the star must be spectrally stable, which means RV is less effective at finding companions in young systems and particularly about A stars (e.g. \citealt{Galland06}). Rotational broadening is also a problem for these stars. However it is in these systems that imaging of outer companions is most successful, due to the decreasing companion luminosity with time \citep{Baraffe03}. For this reason RV and imaging surveys do not usually target the same stars, although there is some overlap between the two techniques. Finally RV cannot detect companions if the system is face on, which is the orientation favoured if the outer companion's motion is to be determined via direct imaging. As a result of these caveats many stars, including almost all of the systems on Figure \ref{fig: a_d_tau_plot}, do not have RV data so large inner masses may not be ruled out. Indeed the detection of an eccentric outer companion could provide motivation for RV follow-up, to investigate whether an unseen inner mass is also present and responsible for this apparent high eccentricity.

Finally, stellar astrometry is also reaching the sensitivities required to detect companions, and this method is most sensitive to face-on orbits so could detect those missed by RV. However this method also requires a baseline longer than the inner mass period. If such precision astrometry is available, the object may remain undetected to $3\sigma$ if

\begin{equation}
\mu \lesssim 2 \times 10^{-2} \left(\frac{a_{\rm i}}{10 \rm AU}\right)^{-1} \frac{d}{50 \rm pc} \frac{\theta_{\rm ast}}{1 \rm mas}
\label{eq: astrometry_detection_limit_simple}
\end{equation}

\noindent where $\theta_{\rm ast}$ is the astrometric accuracy, which is currently of the order 1 mas if many reference stars are available in the same field (e.g. \citealt{Benedict02, Sozzetti05}). Due to the required baseline, it is still possible for Jupiter - brown dwarf mass objects to exist at $\gtrsim 10$ AU and remain undetected by precision astrometry.

The upcoming GAIA mission will bring about a significant improvement in astrometric precision, promising to reach sensitivities of 8 $\mu$as \citep{Casertano08}. However even GAIA will not be able to rule out many massive inner companions, due to the requirement it observes the star for at least one full orbit of the inner body. Using the detection limits from Figures 21 and 22 in \citet{Casertano08} we see that whilst Jupiter mass planets could be detected to three sigma at 2-3 AU from solar type stars out to 200 pc, objects with significantly larger masses could still lie further out as the companion period increases beyond the 5 yr lifetime of the mission. In fact, brown dwarf mass objects could still exist undetected down to 10--20 AU from $1 {\rm M_\odot}$ stars at 10 pc, which means that GAIA will not be able to rule out many of the inner objects required for the scenario in this paper.


\section{Discussion}
\label{sec: discussion}

\noindent We have shown that the orbital elements of an imaged companion may be incorrectly derived due to the presence of an unseen inner mass. We demonstrated that a circular object would always appear eccentric if an inner mass were introduced, and showed that a non-negligible time between observations reduces the effect of this unseen mass on the companion's orbital elements. We then provided a framework to identify the maximum eccentricity error an unseen mass could introduce as a function of readily derivable parameters, and also found the optimum location of such an object. We demonstrated that many imaged companions could potentially be susceptible to this error, and showed that the eccentricity of Fomalhaut-b could be have been overestimated by up to $10\%$. Finally we showed that the large inner masses required by this scenario are not always ruled out by other observations. We will now remark on a few other considerations about this work.

Firstly, we have only examined the effect of an inner mass on the elements of an object that is known to be bound. Another potential application of this scenario is on the initial identification of the companions themselves. If an inner mass lay undetected in a system with a highly eccentric imaged outer object, the stellar velocity shift due to this unseen mass could increase the apparent eccentricity of the observed companion beyond unity. In other words, this effect could actually make a bound companion appear unbound. It is unlikely that such a companion would be classified as a background object, as this would require its apparent motion to mimic that expected of a background source. However one could envisage a scenario where this effect could be important, for example if searching for companions around young stars still in stellar associations. Here an unseen inner mass could lead to a bound imaged object being misidentified as an unassociated member of the same moving group.

As an example we examined the survey of \cite{Janson11}, who imaged 18 massive stars in the solar neighbourhood to search for potential companions. We find three stars amongst their sample, Bellatrix (HIP 25336), Elnath (HIP 25428) and $\lambda$ Aquilae (HIP 93805), which each have point sources located nearby with apparent relative velocities only just large enough to make them unbound. We find that, were unseen inner masses of 50-100 ${\rm M_J}$ located close to these stars (which are below the detection limits for this survey at the required radii), the point sources could be bound (albeit with a high eccentricity). Such sources would be unlikely to be re-imaged in the near future as no companions were identified, so it could be that bound objects are missed. Whilst we consider this unlikely for these three stars due to the large inner masses required, it does highlight the potential importance of this effect for survey work.

Secondly we have only considered two epochs of observation in this paper, which is the minimum number required to estimate the outer body's orbital elements. If more epochs were available, the method presented here would still be applicable so long as the motion of the outer companion appeared linear over the entire observational baseline. In this case the companion positions could be fitted with a linear trend and the problem treated as before. If instead the additional epochs allowed orbital acceleration to be detected then $\tau$ would be too large for an inner body to significantly affect the companion's elements anyway, and so the scenario in this paper would not be applicable.

There is one difference between the use of two and three or more observations however. If an inner object existed with a period less than the time between the first and last observations, the motion of the outer companion should not in fact be linear but should show short term oscillations as the inner binary rotates. This could potentially be identified using the additional observations between the first and last epochs. If such oscillatory motion were detected then this would provide strong evidence for the presence of an unseen companion, and its orbital properties could be constrained. Alternatively if no such motion were observed then it may be possible to rule out significant unseen masses with periods shorter than the observational baseline.

Finally we have assumed that the inner mass is on a circular orbit throughout this work, as this requires fewer parameters than a more general case where both bodies are eccentric. However this need not be the case. As the difference between the astro- and barycentric coordinates of the outer companion increases with the velocity of the star in the barycentric frame, it is clear that this difference may be increased if the inner mass were eccentric and at pericentre when the system was observed. Indeed, the optimum set-up for this scenario would be if both the inner and outer bodies had their orbits aligned (i.e. pericentres in the same direction), and both were near pericentre at the time of the observations. However this exact set-up is unlikely, due to the precise alignment involved. To test this we generated $10^7$ systems with randomised parameters, each consisting of a wide eccentric companion and a coplanar inner mass. Half of the systems had eccentric inner masses, whilst the other half had them on circular orbits. For each system we randomised the mean anomalies of these two bodies and calculated the astrocentric elements of the outer companion. We find that making the inner mass eccentric almost always leads to a lower eccentricity error for the outer body, because it is much rarer for the star to have a velocity shift that exactly opposes the outer companion's motion. Also the optimum case, where both bodies are at pericentre at the time of the observation, is very rare because the bodies do not spend much time around pericentre. The odds of making an observation in this configuration are therefore very low. Hence we conclude that, whilst the eccentricity error on an observed companion may be increased if the inner body is eccentric, in practise the error is almost always reduced in this regime. We have therefore not considered this any further.


\section{Conclusions}
\label{sec: conclusions}

\noindent We have shown that the use of direct imaging to derive the orbital elements of companion planets/brown dwarfs could lead to significant errors if an undetected inner mass is also present. The maximum effect of such a body on the derived eccentricity (and hence semi-major axis) of the observed companion has been quantified for various cases, and we have also identified criteria to determine when this effect may be significant. We provide the reader with a step-by-step method to determine the maximum magnitude of this effect for any system, and apply it to several companions as examples. It appears that many of the currently imaged companions could be susceptible to this scenario when they have orbital motion detected.



\appendix

\section{Min($\mathbf{\mu}$) equation for the $\mathbf{e \neq 0}$ case}
\label{sec: appendix}

\noindent Here we present a simplified equation for the minimum value of $\mu$ required to give an eccentric outer object an astrocentric eccentricity $e'$ if the time between observations is small. This is analogous to the $e=0$ case given by Equation \ref{eq: case1 min mu}, and shows good agreement with the lines on Figure \ref{fig: general case 3}. As for the simpler case the full version of this expression contains high orders of $\mu$, however terms arising from orders greater than two are now less dominant because the equation now includes a first order term inside the square root. Therefore the following formulae (up to second order in $\mu$) are sufficiently accurate without an empirical scaling factor:

\begin{equation}
A \mu^2 + B \mu + C \gtrsim 0,
\label{eq: case3 min mu}
\end{equation}

\noindent where

\[
\begin{split}
A &\equiv \alpha^2 \frac{(1+e)(3+e)}{(1-e)^2} \ \\
	&+ (2e+1) \left( 2 \alpha \frac{1+e}{1-e} \
	+ 4 \sqrt{\alpha} \sqrt{\frac{1+e}{1-e}} \
	+ \frac{4}{\sqrt{\alpha}} \sqrt{1-e^2} \right) \ \\
	&+ \frac{2}{\alpha} (2+3e)(1-e) \
	+ (1+e)^2,
\end{split}
\]

\[
B \equiv 2 e \left(1+e + \alpha \frac{1+e}{1-e} + \frac{2}{\sqrt{\alpha}} \sqrt{1-e^2}\right),
\]

\[
C \equiv e^2 - e'^2,
\]

\noindent which may be solved for $\mu$. Note that as $e \rightarrow 0$, $B \rightarrow 0$ and the solution tends to that of Equation \ref{eq: case1 min mu}, and so in this case the empirical factor $F(e')$ will again be required.

\label{lastpage}

\end{document}